\documentstyle[12pt,epsf]{article}
\newcommand{\be}{\begin{equation}}
\newcommand{\ee}{\end{equation}}
\begin{document}
YITP-00-18 
\hspace{10cm}
\today
\\
\vspace{3cm}
\begin{center}
{\LARGE Once more on the BPS bound for the susy kink
\footnote{NSF grant PHY9722101}} \\  \vspace{2cm}
{Andrei Litvintsev \footnote{e-mail: litvint@insti.physics.sunysb.edu} 
and Peter van Nieuwenhuizen \footnote{e-mail: 
vannieu@insti.physics.sunysb.edu}\\
 { \it C.N.Yang Institute for Theoretical Physics, SUNY at Stony Brook}, \\
{\it Stony Brook, NY 11794 } }
\abstract{\it{
We consider a new momentum cut-off scheme for sums over zero-point energies, 
containing an arbitrary function $f(k)$ which interpolates smoothly 
between the zero-point energies of the modes around the kink and those in flat space. A term 
proportional to $\frac{\partial}{\partial k } f(k)$ 
modifies the result for the one-loop quantum mass $M^{(1)}$ as 
obtained from naive momentum cut-off regularization,  
which now agrees with previous results, 
both for the nonsusy and susy case.
We also introduce a new regularization scheme for the evaluation of the 
one-loop correction to the
central
charge $Z^{(1)}$, with a cut-off $K$
for the Dirac delta function in the canonical commutation relations and
a cut-off $\Lambda$ for the loop momentum. The result for $Z^{(1)}$ depends only on 
whether $K>\Lambda$ or $K<\Lambda$ or 
$K=\Lambda$. The last case yields the correct result and saturates the
BPS bound, $M^{(1)}=Z^{(1)}$,
in agreement with the fact that multiplet shortening does occur in this $N=(1,1)$ system.
We show how to apply
mode number regularization by considering first a kink-antikink
system, and also obtain the correct result with this method.
Finally we discuss the relation of these new schemes to previous 
approaches based on the Born expansion of phase shifts
and higher-derivative regularization.
} } 
\end{center}
\newpage

\section{Introduction.}
In the past few years 
several articles [1-8] have
dealt with the issue of the 
one-loop \footnote{Only the 2-loop corrections $M^{(2)}$ in the bosonic sector 
of the sine-Gordon model have been explicitly calculated [9,2].} 
quantum 
corrections $M^{(1)}$ to the mass of ordinary and supersymmetric (susy) solitons in (1+1) dimensions,
and the one-loop quantum corrections
$Z^{(1)}$ to the central charge of  
supersymmetric \footnote{By susy we mean N=(1,1) supersymmetry in (1+1)
dimensions. For the case of N=(2,2) susy in (1+1) dimensions it was 
shown in \cite{misha} that $M^{(1)}=0$ and in \cite{shifman} that $Z^{(1)}=0$.}
solitons in (1+1) dimension. Although this subject was intensively 
studied in the 1970's, different answers for $M^{(1)}$ and $Z^{(1)}$ were 
given at that time in the literature, often without discussing 
regularization and/or renormalization issues. 
No consensus was reached about the final correct values.

With the advent of dualities between extended objects and point-like
objects, these issues were brought back into focus in \cite{pvn}.
Three different approaches were subsequently developed and used to compute $M^{(1)}$ and
$Z^{(1)}$: the Casimir approach of summing over zero-point energies [10,1,2],
an approach based on the Born expansion of phase shifts [3,4,5,7], and an 
approach which recognized that the susy generators are composite operators which
require regularization, leading to an anomaly
\footnote{
The possibility that a topological anomaly might be present was first suggested in \cite{misha} (in the introduction and the conclusions).
The authors of \cite{shifman} were unaware of this suggestion \cite{pismo},
but went much further: the anomaly was identified and a thorough
analysis was carried out in \cite{shifman}.
}
in the central charge 
which restores
the BPS bound $M^{(1)}=Z^{(1)}$ \cite{shifman}. All these approaches now agree
on the correct value of the ordinary (bosonic) kink and the supersymmetric 
kink
\be
M^{(1)}(bos) =  -m \hbar \left( \frac{3}{2 \pi} -
\frac{\sqrt{3}}{12} \right); \hspace{1cm}
M^{(1)}(susy) = Z^{(1)}(susy) = - \frac{\hbar m}{2 \pi} 
\label{ura1}
\ee

The correct value for $M^{(1)}$ 
in the bosonic case was first obtained in \cite{dashen}, while
the correct value for $M^{(1)}(susy)$ was first obtained in 
\cite{schonfeld}, but, as noted in \cite{schonfeld}, there
were ambiguities in the derivation. The first consistent 
and unambiguous calculation of the correct values for 
$M^{(1)}(bos)$ and $M^{(1)}(susy)$ was carried out in \cite{misha}.
The correct values for $Z^{(1)}$ were first obtained in \cite{graham},
but the derivation has been criticized in \cite{shifman} since it
nowhere used the concept of anomalies, and another
derivation based on anomalies was given in \cite{shifman}.

The reasons we nevertheless come back to this problem are twofold. On the
one hand, we (and others) believe that there are still open questions in the
existing derivations. This is not surprising: the field has been confusing
for decades
and deals with very subtle issues concerning renormalizability conditions of
quantum field theories in a nontrivial background. On the other hand,
we have constructed two new derivations , one for $M^{(1)}$ and one for $Z^{(1)}$,
which resolve some questions and shed new light on previous derivations. 

It has recently been claimed \cite{shifman} that in $N=(1,1)$
susy models in $1+1$ dimensions multiplet shortening does occur.
There are some aspects of this problem having to do with
$Z_2$ gauge symmetry due to the fermionic zero modes which we shall study 
elsewhere \cite{fred}, but we believe that multiplet shortening indeed does
occur, hence 
$M=Z$ to all loop orders, and the various calculations of
$M$ and $Z$ must pass the
consistency check that
$M=Z$.

The new derivation for $M^{(1)}$ in section 2 is 
based on the quantization condition $k_n L + \delta(k_n) 
f(k_n)= 2 \pi n + a$ instead of
$k_n L + \delta(k_n) = 2 \pi n + a$ (with $a=0$ or $a=\pi$)
for bosons and a similar one for fermions.
Here $f(k)$ is a smooth 
(or non-smooth) cut-off function which 
gives extra contributions to $M^{(1)}$.
In the density of states $\frac{dn}{dk}$ there is one modification which is very 
natural, namely the replacement of $\delta^\prime(k_n) $ by $\delta^\prime(k_n) f(k_n)$,
but there is a second modification, namely a term $\delta(k_n) f^\prime(k_n)$, which is new, 
and crucial to obtain the correct result.

With this extra contribution also the momentum cut-off scheme now gives the
result in (\ref{ura1}). We shall discuss
this quantization condition in detail in section 2,
but we note here that this approach is very natural for the ordinary 
Casimir effect. Namely, modes with sufficiently high energy leak through the 
plates and become equal to the modes without plates, initially only a bit 
($f 
\sim 1$) but more and more so as their energy increases ($f \sim 0 $).

The problem with quantum corrections to the central charge $Z$ is that it is classically a total derivative and this seems to exclude quantum corrections \cite{pvn}.
However, after regularization of $Z$ it ceases to be a total derivative, and ordinary loop corrections occur.
The new derivation for $Z^{(1)}$ in section 3 uses a very general approach with
two rather than one cut-offs: one cut-off ($K$) for the Dirac delta functions 
which appear in the evaluation of the equal-time canonical commutation relations used to
evaluate the susy algebra, and another cut-off ($\Lambda$) used to regulate the momentum of the one-loop graphs contributing to $Z^{(1)}$. Depending on 
$K>\Lambda$, $K<\Lambda$ or $K=\Lambda$ one retrieves results of previous approaches.
On physical grounds, we claim that one should use $K=\Lambda$, and this indeed 
yields (\ref{ura1}).
 
In section 4 we comment on some unsolved questions in previous approaches 
(including our own). 
It has been observed that the result $M^{(1)}= Z^{(1)}$ has been
obtained in the phase shift approach without apparently encountering an anomaly
in $Z^{(1)}$. Also the issue whether the phase shift
approach is based implicitly on a particular  
regularization scheme will be discussed. 
Another issue we study is whether higher derivative regularization really 
regulates the graphs for $Z^{(1)}$. 
(In gauge theories it is well-known that higher-derivative regularization is by itself not sufficient to
regulate one-loop graphs).
In the Casimir approach, the mode cut-off scheme gave 
a correct result for $M^{(1)}$ in the bosonic case but not in the susy case \cite{pvn}.
Instead a scheme was developed based on first computing 
$\frac{\partial}{\partial m}M^{(1)}$ (which is less divergent) \cite{misha}. This gives
the correct result whenever the boundary conditions are such that the divergencies in $M^{(1)}$ cancel, but the question remained why mode  
cut-off did not work in the susy case. 
We shall show that mode cut-off gives the correct result in (\ref{ura1})
provided one counts the number of 
bosonic and fermionic zero modes correctly.
Furthermore we have been asked why in the
$\frac{\partial}{\partial m}M^{(1)}$ scheme we only differentiate the energies
$\sqrt{k^2+m^2}$ but not the densities $\rho(k)$ of the states.
Also this question will be answered.

We shall discuss these issues in section 4 but we stress that no doubt exists
on our part about the correctness of (\ref{ura1}). Furthermore, each scheme has
its own beauty, and they are all based on physical principles which are 
likely to play a role in the future. However, since quantum field theory
with both point-like and extended objects is undoubtedly going to become an
area of active research, all subtle renormalization and regularization
issues in the derivations of (\ref{ura1}) should be totally clear.   

In the rest of this introduction we review some basics.
The Lagrangian for the N=(1,1) susy kink is given by
$
{\cal L}= -\frac{1}{2} (\partial_\mu \phi_0)^2 - \frac{1}{2} U^2(\phi_0)
- \frac{1}{2} \bar{\psi_0} \gamma^\mu \partial_\mu \psi_0 - \frac{1}{2}
\frac{dU}{d \phi} \bar{\psi_0} \psi_0
$ where $U[\phi_0(x)] = \sqrt{\frac{\lambda_0}{2}} \left( \phi_0^2 - 
\frac{\mu_0^2}{\lambda_0} \right) $ and $\bar{\psi_0}=\psi_0^{\dagger} 
i \gamma^0$
with $(\gamma^0)^2= -1$.
The susy transformation rules, after eliminating the auxiliary field $F$,
are $\delta \phi= \bar{\epsilon} \psi$ and $\delta \psi = \not{\partial}
\phi \epsilon - U \epsilon$.
The theory is renormalized by replacing $\mu_0^2$ by $\mu^2+\delta \mu^2$
where $\delta \mu^2$ is a counter term. 
The precise definition and value of $\delta \mu^2$ is one of the main
problems to be discussed.
No wave
function and coupling constant renormalization is needed in 1+1
dimensions, hence we set $\phi_0=\phi$, $\psi_0=\psi$  and 
$\lambda_0=\lambda$, where $\phi$, $\psi$ and $\lambda$ are renormalized
objects.  
This amounts to a particular choice of renormalization conditions 
\cite{pvn}.
Expanding $\phi$ about the
solution $\phi_{cl}=\frac{\mu}{\sqrt{\lambda}}$ as $\phi=\phi_{cl}+\eta$
one obtains
$$
{\cal L}= -\frac{1}{2} (\partial_\mu \eta)^2 - \mu^2 \eta^2 -
\mu \sqrt{\lambda} \eta^3 - \frac{1}{4} \lambda \eta^4 + \frac{1}{2}
(\delta \mu^2)_{susy} \left( \eta^2+ 2 \frac{\mu}{\sqrt{\lambda}} \eta \right) -
\frac{1}{4} \frac{(\delta \mu^2)^2_{susy}}{\lambda}
$$
\be
-\frac{1}{2} \bar{\psi} \gamma^\mu \partial_\mu \psi 
- \frac{1}{2} ( \sqrt{2} \mu  ) \bar{\psi} \psi 
- \sqrt{\frac{\lambda}{2}} \eta \bar{\psi} \psi
\ee
Hence the renormalized meson mass parameter is $m^2 \equiv 2 \mu^2$, and the  
fermion has the same mass, as expected from susy.

The mass 
counter term $\delta \mu^2$ is fixed by requiring that it cancels the tadpole
graphs 
\be 
\left( \delta \mu^2 \right)_{susy} = 
\frac{\lambda \hbar}{4 \pi} \int_{-\Lambda}^{\Lambda}
\frac{dk}{\sqrt{k^2+m^2}}; \hspace{0.5cm}
\epsfxsize 1.3in \raisebox{-0.4cm}{\epsfbox{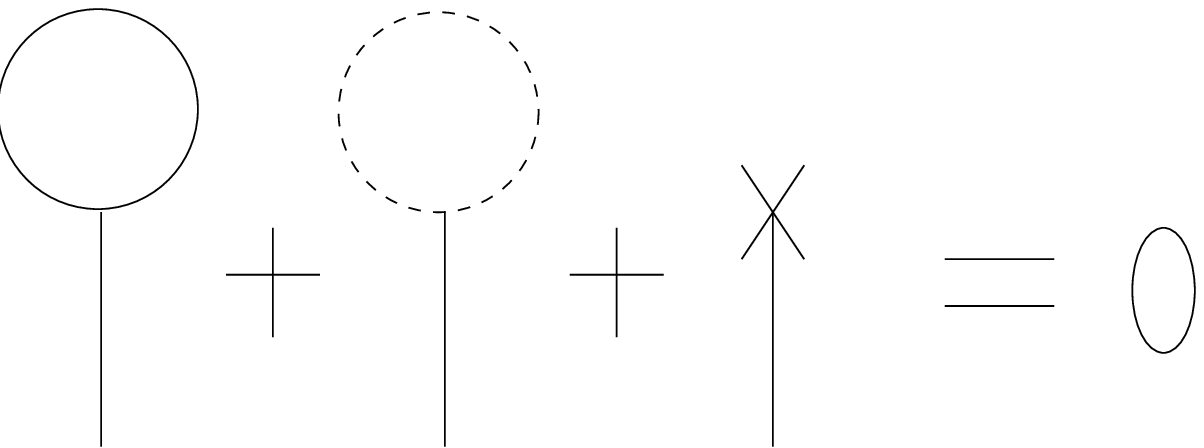}}
\ee
The bosonic loop (denoted by a solid circle) gives a result proportional
to $\frac{3 \lambda \hbar}{4 \pi}$ and the fermionic loop 
(denoted by a dotted circle) is proportional to $-\frac{\lambda \hbar}{2 \pi}$.
Further $\Lambda$ is a momentum cut-off; the same cut-off
should be used in the one-loop graphs such that
the sum of graphs in (3) cancels at the regularized level without
leaving any finite part. 
This is the renormalization condition of [1] which was also adopted 
in [2,3,4,5]. 

Having fixed $(\delta \mu^2)_{susy}$ 
once and for all, we expand about the kink by setting 
$\phi=\phi_K(x)+\eta(x,t) $ where $
\phi_K(x)=\frac{\mu}{\sqrt{\lambda}} \tanh \frac{\mu x}{\sqrt{2}} $ is the
classical kink solution, 
with mass $M^{(0)}=\frac{m^3}{3 \lambda}$,
and $\eta$ is the meson field of fluctuation around
the kink. 
The background satisfies the Bogomol'nyi equation 
$\partial_x \phi_K+U(\phi_K)=0$ and 
is invariant under susy transformations with parameter 
$\frac{1}{2}(1-\gamma_1) \epsilon$.
The one-loop action  then reads 
$$
{\cal L}= -\frac{1}{2} (\partial_\mu \eta)^2 - \frac{1}{2} m^2 \eta^2 
- \frac{3}{2} \lambda \left( \phi^2_K -\frac{\mu^2}{\lambda} \right) \eta^2
+ \frac{1}{2} (\delta \mu^2)_{susy}  \left(
\phi_K^2- \frac{\mu^2}{\lambda} \right) 
$$
\be
\label{f2}
-\frac{1}{2} \bar{\psi} \gamma^\mu \partial_\mu \psi 
-\frac{1}{2} m \bar{\psi} \psi 
- \sqrt{\frac{\lambda}{2}}
\left( \phi_K - \frac{\mu}{\sqrt{\lambda}}  \right) \bar{\psi} \psi
.
\ee
We shall be careful and explicit about which parts of loop graphs in the topological
sector will be canceled by the counterterm in (\ref{f2}).

The quantum Hamiltonian $\hat{H}_{qu}$ depends on the field $\hat{\eta}$ and 
its conjugate momentum $\hat{\pi}$, and on the collective momentum $\hat{P}$,
but it is independent 
of the collective coordinate $\hat{X}$ for the center of mass. 
The quantum mass of the 
kink is the vacuum expectation value of $\hat{H}_{qu}$. For a kink at rest 
$\langle \hat{P} \rangle$
vanishes, and the one-loop quantum Hamiltonian reduces to
\be
H=\int_{-\infty}^{\infty} \left[ \frac{1}{2} \pi^2 + \frac{1}{2} \left(
\frac{\partial}{\partial x} \eta 
\right)^2 + \frac{1}{2} \eta^2 (U U^{\prime \prime}+U^\prime U^\prime) (\phi_K)
+ \frac{1}{2} \bar{\psi} \gamma^1 \partial_x \psi 
+ \frac{1}{2} U^\prime (\phi_K) \bar{\psi} \psi 
 \right] dx  + \Delta M_{susy}
\label{hn}
\ee
where $U^\prime = \frac{dU}{d \phi}$, 
$U^{\prime \prime}= \frac{d^2U}{d \phi^2}$ and 
$\Delta M_{susy}$ is the mass counter term for the sector with the susy kink, 
given according to (\ref{f2}) by
$\Delta M_{susy} = - \Delta L_{susy} =  \frac{m}{\lambda} (\delta
\mu^2)_{susy} $.

The standard formula for the one-loop correction to the mass of the 
bosonic kink follows from (\ref{hn}) and reads \cite{raj}
\be
M^{(1)}=\sum \frac{1}{2} \hbar \omega_n - \sum \frac{1}{2} \hbar \omega_n^{(0)}
+ \Delta M_{bos}
\label{f3}
\ee
Here the first term gives the sum of the zero point energies of bosons
(including the bound state and the normalizable zero mode) 
in the presence of the kink, while the second
term gives the same quantities in flat space.  
The mass counterterm is still given by $\Delta M_{bos}=\frac{m}{\lambda} 
\delta \mu^2$ but in the bosonic case 
$(\delta \mu^2)_{bos}= 3 \lambda \hbar / 4 \pi \int_{-\Lambda}^{\Lambda}
dk / \sqrt{k^2+m^2} $.
All terms in (\ref{f3}) are divergent and depend on $m$ but not on the coupling constant $\lambda$.
The difference of the two series is only logarithmically divergent, and adding $\Delta M_{bos}$ one
obtains for certain boundary conditions a finite answer, but the crucial question is then how to determine
unambiguously the finite answer. 

For large $x$, the bosonic and fermionic quantum
fluctuations with a continuous spectrum 
in a representation with $\gamma^1= \left( \begin{array}{cc} 
1 & 0 \\ 0 & -1 
\end{array}
\right)$
and
$\gamma^0= \left( \begin{array}{cc} 
0 & -1 \\ 1 & 0  
\end{array}
\right)$
read 
$\eta(x,t) \to e^{i \left( kx-\omega t \pm \frac{1}{2} \delta(k) \right)}$;
$\psi_1(x,t) \to e^{i \left( kx-\omega t \pm \frac{1}{2} \delta(k) \right)}$;
$\psi_2(x,t) \to e^{i \left( kx-\omega t \pm \frac{1}{2} \delta(k)  \pm
\frac{1}{2} \theta(k) \right)}$ for $x \to \pm \infty$. 
The phase shifts are sketched in figure 3.
The bosonic phase shift is given by $\delta(k)  = \left(
2 \pi - 2 \arctan \frac{3 m |k|}{m^2-2k^2}
\right) {\rm sign} (k)
$
and only depends on $k$ and $m$ through $k/m$. 
It tends to zero for $|k| \to \infty$ and jumps at $k=0$ from $- 2 \pi$ to
$2 \pi$.
The phase shift $\theta(k)$ is given by $\theta(k)=-2 \arctan m/k$ and
tends also to zero for large $|k|$, but it jumps from $+\pi$ to $-\pi$
at $k=0$.   
Imposing boundary conditions quantizes the
momenta $k$ but different boundary conditions will lead in general to
different quantization rules.
Taking anti-periodic boundary conditions on the quantum fluctuations
$\eta$ about a kink in a box of length $L$, 
the momenta are quantized by 
\be
k_nL+\delta(k_n) = 2 \pi n + \pi
\label{h1}
\ee
The bosonic density of states in the presence of the kink and in the trivial sector 
 are given by
\be
\label{mm1}
\frac{dn}{dk}=\rho(k) = \frac{ L}{2 \pi} \left(1+\frac{1}{L} 
\frac{\partial}{\partial k}
\delta \left({k} \right) \right)
;  \hspace{0.5cm} 
\rho_0(k) = { L}/{2 \pi} .
\ee

In the bosonic case one can 
also impose periodic boundary conditions \cite{pvn}, but in the susy 
case periodic boundary conditions are ruled out for the fermions
if one uses plane waves (see eq (32) and eq (17) of \cite{misha}). This follows from the consistency
of the equation
\be
\left( \begin{array}{c} e^{-ikL/2-\frac{1}{2} \delta(k)} \\
 e^{-ikL/2-\frac{1}{2} \delta(k)- \frac{1}{2} \theta(k) } \\ \end{array}
\right) =
\Gamma
\left( \begin{array}{c} e^{ikL/2+\frac{1}{2} \delta(k)} \\
 e^{+ikL/2+\frac{1}{2} \delta(k)+ \frac{1}{2} \theta(k) } \\ \end{array}
\right)
\label{granferm}
\ee
where $\Gamma = \left( \begin{array}{cc} 1 & 0 \\ 0 & 1 \\ \end{array} \right) $
for periodic boundary conditions is inconsistent. The twisted boundary conditions with
$\Gamma = \left( \begin{array}{cc} 0 & 1 \\ 1 & 0 \\ \end{array} \right) $
or 
$\Gamma = \left( \begin{array}{cc} 0 & -1 \\ -1 & 0 \\ \end{array} \right) $
are consistent.
We stress that this does not mean that one can not use 
periodic boundary conditions for fermions at all. It only means
that one should use standing waves instead of plane waves, as shown in \cite{pvn}.

Other boundary conditions in the kink sector which we shall use are the 
supersymmetric boundary conditions $\eta=\psi_1=0$ at $x=\pm L/2$ of \cite{schonfeld} and another set of supersymmetric 
boundary conditions $\partial_x \phi +U = \psi_2=0$ at $x=\pm L/2$ given in \cite{shifman}\footnote{
The background is invariant under susy transformation with  $\epsilon_2=
\frac{1}{2} (1-\gamma^1) \epsilon$, while under linearized $\epsilon_2$ susy transformation these linearized boundary conditions
transform into each other. For example, $\delta \psi_2 $
transforms into
$ -( \partial_x \eta \pm m) \epsilon_2 $ at $x=\pm L/2$, which is indeed the linearized part 
of $\partial_x \phi +U=0$. In ref. \cite{shifman} also a condition $(\partial_x + U^\prime) \psi_1 = 0 $ is mentioned, which 
follows from $\psi_2=0$ and the Dirac equation. 
}.

\section{A new method for momentum cut-off regularization of $M^{(1)}$.}

In ref \cite{pvn} it was discovered that the widely used momentum cut-off regularization
for the sums in (\ref{f3}) gave different results from the equally widely used mode number 
cut-off. (We refer to \cite{pvn} for earlier references).
This puzzling difference renewed research
activity in this area. 
In \cite{misha} another approach was developed to compute $M^{(1)}$, namely by first evaluating the 
sums of the derivatives $\frac{\partial}{\partial m} \omega_n $
and  $\frac{\partial}{\partial m} \omega_n^0 $ and then integrating w.r.t.
$m$. This gave
the correct result (for those cases where the divergencies in
$M^{(1)}$ cancel, see section (4.1) ). We shall call this derivative regularization.
However, the $\frac{\partial}{\partial m} M^{(1)}$ approach, though 
successful, replaced the momentum cut-off approach, but did not 
pinpoint the place where the momentum cut-off scheme went wrong.
Here we present a modified version of the momentum cut-off method. 
As we have 
already mentioned in the introduction, the results based on the new 
momentum cut-off agree with the generally accepted 
expressions in (\ref{ura1}). 

We now make a careful analysis of the various steps which constitute the
momentum cut-off scheme.
A natural regularization 
procedure would be to disregard 
modes of the quantum fields with momenta $k$  
above some ultraviolet cut-off $\Lambda$. 
We did already use this regularization
for our counterterm, which has a 
logarithmic ultraviolet divergence.
However, the sums over zero-point energies are linearly divergent,
and in such cases it makes a difference how we cut-off the momenta.
It may help to identify the momentum cut-off regularization used here with 
the plasma frequency used in the usual Casimir effect. Then modes with
energies much lower than the plasma frequency are completely defined
by the potential profile, modes with the energy of the order of the plasma
frequency start to leak through the potential, and finally, modes with
energies much higher than the plasma frequency do not notice the presence of
the potential at all.

Consider the quantization condition with a sharp cut-off for the modes
around the bosonic kink. We begin with anti-periodic boundary conditions 
both in the nontrivial sector with the kink and in the trivial sector. Then
\be
\label{ss0}
k_n L + \delta(k_n) \Theta (\Lambda-k_n)= 2 \pi n + \pi
\ee
\be
\label{qs0}
k_n^0L=2 \pi n + \pi
\ee
where $\Theta(\Lambda-k_n)=1$ for $k_n<\Lambda$, 
$\Theta(\Lambda-k_n)=0$ for $k_n>\Lambda$ and $0 \le \Theta(0) \le 1$.
We shall later consider smooth versions of this theta function.
We think of $n=n(k)$ as the function counting the number of states in the 
descretized version of the continuum spectrum. The spectral density 
in the non-trivial sector is 
then given by
\be
\label{den1}
\frac{dn}{dk} = \frac{1}{2 \pi} 
\left[ L -\delta_{Dirac}(k-\Lambda) 
\delta(k) +
\Theta(\Lambda-k) \frac{d \delta(k)}{dk}  \right] 
\ee
Notice that it does not matter whether one uses antiperiodic or periodic boundary conditions
(as long as they are the same in the nontrivial sector and the trivial sector)
since
the constant $\pi$ drops out in the differentiation. This independence of
$M^{(1)}$ from certain sets of
boundary conditions for the bosonic kink was already observed in \cite{misha}.
One may rewrite (\ref{f3}) then as 
\be
\label{lena}
M^{(1)} =\frac{\sqrt{3} \hbar m}{4} +
\frac{1}{2} \hbar \sum \sqrt{k_n^2+m^2} - \frac{1}{2} \hbar \sum \sqrt{(k_n^{(0)})^2+m^2} + 
\frac{3 \hbar m}{2 \pi} \int_0^\Lambda \frac{dk}{\sqrt{k^2+m^2}}
\ee
where the first term is the contribution from the bound state and the last
term is the counter term. 
We replace the sums by integrals
\be
M^{(1)} = \frac{\sqrt{3} \hbar m}{4}+
\hbar \int_0^\infty \frac{dk}{2 \pi} \left( \frac{dn}{dk} - 
\frac{dn^{(0)}}{dk} \right) \sqrt{k^2+m^2} 
+ \frac{3 \hbar m}{2 \pi} \int_0^\Lambda \frac{dk}{\sqrt{k^2+m^2}} 
\ee
Using (\ref{den1}) and $dn^{(0)}/dk = L/2 \pi$ we obtain 
$$
M^{(1)} = \frac{\sqrt{3} \hbar m}{4}+
\hbar \int_0^\infty \frac{dk}{2 \pi} 
\Theta(\Lambda-k) \frac{d \delta(k)}{dk}
\sqrt{k^2+m^2} -
\hbar \int_0^\infty  
\frac{dk}{2 \pi} \delta_{Dirac}(k-\Lambda) \delta(k) \sqrt{k^2+m^2}  
$$
\be
+ \frac{3 \hbar m}{2 \pi} \int_0^\Lambda \frac{dk}{\sqrt{k^2+m^2}}. 
\ee
For fixed $\Lambda$ all terms are finite, but for $\Lambda \to \infty$ 
the counter term combines with the first integral, rendering
the sum finite.  
The second integral is the new term in $M^{(1)}$. It has a finite limit as 
$\Lambda \to \infty$, and an easy 
computation reproduces the result (\ref{ura1}). 
 
Notice that the difference from the naive momentum cut-off comes from the term
$  M^{(1)}- M^{(1)}_{naive} = - \hbar \int_0^\infty  
\frac{dk}{2 \pi} \delta_{Dirac}(k-\Lambda) \delta(k) \sqrt{k^2+m^2} $.
It seems to depend crucially on the Dirac delta function. 
What happens if we use a smooth cut-off function $f(k)$ instead
of $\Theta(\Lambda-k)$, such that it is still 
equal to zero for $k> \Lambda+\Delta \Lambda$ and equal to one for 
$k< \Lambda-\Delta \Lambda$, while it interpolates smoothly between
these two values for $k$ in $(\Lambda-\Delta \Lambda,\Lambda+\Delta \Lambda )$?
Then 
\be
M^{(1)}- M^{(1)}_{naive} =  \hbar \int_0^\infty  
\frac{dk}{2 \pi} \frac{df}{dk} \delta(k) \sqrt{k^2+m^2} 
\ee
and the integral is localized around $\Lambda$. For large $\Lambda$ we can 
write 
\be  
\label{s17}
M^{(1)}- M^{(1)}_{naive} = \frac{1}{2 \pi} \hbar \left( \int_0^\infty  
{dk} \frac{df}{dk}  \right) \delta(\Lambda) \sqrt{\Lambda^2+m^2} 
\ee
and the integral in the above expression is equal 
to minus unity for any smooth $f(k)$. 
Thus we observe that our result is not sensitive to the exact shape of $f(k)$.

Next we consider the case that the bosonic fluctuations are anti-periodic
in the kink sector, but periodic in the trivial sector. The quantization conditions become 
\be
k_nL + [\delta(k_n)-\pi] \Theta(\Lambda-k_n)=2 \pi n; \hspace{0.5cm} 
k_n^0L=2 \pi n
\ee
Note that in the kink sector we smoothly deform the
anti-periodic boundary conditions at lower energy
into the periodic boundary conditions of the trivial sector.
One obtains now an extra term in $dn/dk$, namely $\delta_{Dirac}(\Lambda-k) \pi
$, which raises $M^{(1)}$ by the infinite amount 
$\frac{1}{2} \hbar \sqrt{\Lambda^2+m^2}$.
One should thus use the same boundary conditions (periodic or anti-periodic) 
for the bosonic case 
in both sectors; otherwise the result for $M^{(1)}$ diverges. 
The physical origin of this infinite energy was already
pointed out in \cite{schonfeld}: one needs extra energy to change
the boundary conditions in the trivial sector to other boundary 
conditions in the nontrivial sector. This extra energy should not be
part of the mass of the kink.

We now turn to the supersymmetric case. 
First of all, we require that in the trivial sector the contributions
from the bosons and fermions cancel. For example 
the sum of the bosonic and fermionic zero modes
given by (\ref{qs0}) and (\ref{qs1}) cancels. We may also 
use periodic boundary conditions for the bosons and fermions in the trivial sector, or any other
boundary condition in the trivial sector, as long as their sum cancels.
In the computation of $M^{(1)}$ it is not important whether one uses periodic or antiperiodic 
boundary conditions because
an extra constant term $\pi$ in the quantization condition for bosons
and/or fermions cancels in $dn/dk$. It only matters that the leading term 
in the quantization conditions is $2 \pi n$. Thus, for example, also the
boundary conditions $\eta(-L/2)=\eta(L/2)=0$ and $\psi_2(-L/2)=\psi_2(L/2)=0$ give the same correct result \footnote{
Of course these are not susy boundary conditions, because as we have discussed, the latter give an infinite value for $M^{(1)}$. 
}.

Consider the case of twisted boundary conditions for fermions in the kink sector.
We use our regularization prescription to rewrite the twisted antiperiodic boundary 
conditions for fermions of \cite{misha} (corresponding to (\ref{granferm}) with
$\Gamma = \left( \begin{array}{cc} 0 & -1 \\ -1 & 0 \\ \end{array} \right) $ ) in the form
\be
\label{ss1}
k_nL+\Theta(\Lambda-k_n) \left( \delta(k_n) +\frac{\theta(k_n)}{2} \right) = 2 \pi n + \pi  
\ee
\be
\label{qs1}
k_n^0L = 2 \pi n + \pi
\ee
where $\theta(k)= - 2 \arctan \frac{m}{k} $. For the bosons we use the antiperiodic boundary conditions in (\ref{ss0}).
We then obtain 
\be
M^{(1)}(susy)=\left( \sum \frac{1}{2} \hbar \omega^B_n - \sum \frac{1}{2} \hbar \omega_n^{B(0)}
\right) - \left( \sum \frac{1}{2} \hbar \omega^F_n - \sum \frac{1}{2} \hbar \omega_n^{F(0)}
\right)
+ \Delta M_{susy}
\label{fuf3}
\ee
using exactly the same steps as earlier in this section. 
In the difference the contributions with $\delta(k)$ cancel [1,2,6]. 
Furthermore, the naive momentum cut-off approach gives zero 
(see eq. (59) of \cite{pvn}). Thus, only the extra term remains, and with 
(\ref{fuf3}) and (\ref{ss1}) one gets
\be 
M^{(1)}(susy)=
\frac{1}{2} \hbar \int_0^\infty  \frac{dk}{2 \pi} \theta(k) \sqrt{k^2+m^2} \delta_{Dirac}(k-\Lambda).
\label{ps22}
\ee
This yields the  
expression (\ref{ura1}) for the correction to the mass of the 
supersymmetric kink.
As before, the result holds for any smooth cut-off function. 

One obtains the same (correct) result if one drops the term $+ \pi$
in (\ref{ss1}),(\ref{qs1}) and/or (\ref{ss0}),(\ref{qs0}) because
in $\frac{dn}{dk}$ such terms clearly cancel. Thus for the fermions one may use twisted periodic or twisted antiperiodic boundary conditions or their average (one-half the sum), while for the bosons one may use periodic or antiperiodic boundary conditions or their average.

In the kink sector some sets of boundary conditions clearly give incorrect result. For
example , for supersymmetric boundary conditions the bosonic zero-point energies cancel the fermionic ones, so that one is 
only left with the divergent counterterm. We shall later consider the kink-antikink system and derive from this
study correct sets of boundary conditions for a single kink. One correct set which we shall in this way obtain is the average of
twisted periodic and twisted antiperiodic conditions for the
fermions, and the average of periodic and antiperiodic conditions for the bosons.   

We conclude that momentum cut-off regularization is rehabilitated
as a regularization scheme which yields correct results in all cases,
but one should not forget the cut-off functions in (\ref{ss0}) 
and (\ref{ss1}).

\section{A new momentum cut-off regularization for the central charge.}

The possibility that a topological quantum anomaly plays
a role in the one-loop corrections to $M^{(1)}$ and $Z^{(1)}$ was first
mentioned in \cite{misha} (in the introduction and conclusion), 
but the full
appreciation of the role of an anomaly in the central charge 
based on supermultiplets and on explicit calculations in several regularization schemes  
was given in 
\cite{shifman}.
We shall comment on higher-derivative regularization
employed in \cite{shifman} in section 4.
In this section we give a derivation of the central charge 
anomaly with a very simple scheme. 

We make the regularization a little bit more general than one would need 
in order to be able to compare this scheme with the higher-derivative scheme of
\cite{shifman}. We impose
two different cut-offs. 
We define the canonical commutation relations  
\be \left[ \phi(t,x),\partial_0 \phi(t,x^\prime) \right] =
i \hat{\delta}(x-x^\prime) ;
\hspace{1cm}  
\left\{\psi^\alpha(t,x),\psi^{T\beta}(t,x^\prime)\right\}=\delta^{\alpha \beta}
\hat{\delta}(x-x^\prime)
\ee
(For Majorana spinors with a real representation for the Dirac matrices, 
$(\psi^\beta )^* \equiv \psi^*_\beta$ is equal to $\psi^\beta$). 
We keep a finite number of Fourier modes
in the delta function, i.e.
\be
\hat{\delta}(x) = \int_{-K}^K \frac{dq}{2\pi} \exp(iqx)
\label{dp1}
\ee
We also 
keep only a finite number of Fourier modes 
in the propagators
\be
\label{dp2}
\langle \phi(x) \phi(y) \rangle = \int_{-\Lambda}^\Lambda \frac{dp}{4 \pi}
\frac{\exp(ip(x-y))}{\sqrt{p^2+m^2}} 
\ee
We could have used
distorted plane waves in (\ref{dp1}) and (\ref{dp2}), namely solutions of the field equations for small
fluctuations around the kink. In that case our cut-off procedure would be a cut-off in the number of modes. Consistency of the procedure would require $\Lambda=K$.
We shall prove (see (\ref{prv1}) ) that the computation with distorted plane waves gives
exactly the same result as with plane waves because the difference is a total derivative which vanishes.  

The classical Noether current for supersymmetry is 
\begin{equation}
j^\mu = -( \partial_\nu \phi) \gamma^\nu \gamma^\mu  \psi - U 
\gamma^\mu \psi
\end{equation}
(In section 4 we discuss extra terms in $j^\mu$ of the form
$\epsilon^{\mu \nu} \partial_\nu X$).
We compute the anticommutator of the supercharges 
\be
\{ Q^\alpha, \bar{Q}_\beta \} = 2 i (\gamma^\nu)^\alpha_{\hspace{1.5mm}\beta} 
P_\nu- 2i 
(\gamma^5)^\alpha_{\hspace{1.5mm} \beta} Z
\ee 
\be
Q^\alpha = \left( \begin{array}{c} Q_1 \\ Q_2 \\ \end{array} \right);
\{ Q_1, Q_1 \} = 2 (H+Z); \{ Q_2,Q_2 \} = 2 (H-Z); \{ Q_1,Q_2 \} = 2 P 
\ee
The one-loop central charge 
contributions come from the term proportional to $\gamma^5=\gamma^0 \gamma^1 $
which results from two sources: \\
1) from taking the anticommutator 
of the fermionic field in the first term in the current 
with the fermionic field
in the second term in the current, and then expanding to second
order in bosonic fluctuations and computing the bosonic loop 
which contributes to the vacuum expectation value;\\
2) from the commutator of the $\partial_0 \phi$ term in the first term of the
current with the bosonic field $\phi$ in the second term of the current, and then
computing the fermionic loop which contributes to the vacuum expectation 
value.

For the case 1), we obtain
\be 
Z_{bos}=\int_{-L/2}^{L/2} dx \int_{-L/2}^{L/2} dx^\prime U[\phi(x)]
 \partial_{x^\prime} \phi(x^\prime) \hat{\delta}(x-x^\prime)
\ee
where we have put the system in a box of length $L$ to regulate 
infrared divergences. 
Expanding $\phi(x,t)=\phi_K(x)+\eta(x,t)$ and 
computing the vacuum expectation value up to one loop, we get
$$ \langle Z_{bos} \rangle =Z_{cl}+\frac{1}{2} \int_{-L/2}^{L/2} dx 
\int_{-L/2}^{L/2} dx^\prime
U^{ \prime \prime}[\phi_K(x)]
\partial_{x^\prime} \phi_K(x^\prime)
\langle
\eta^2(x) \rangle
\hat{\delta}(x-x^\prime)   
$$ 
\be
\label{bz}
 + \int_{-L}^{L} dx \int_{-L}^{L} dx^\prime 
U^{ \prime}[\phi_K(x)]
\langle \eta(x) \partial_{x^\prime} \eta(x^\prime) \rangle 
\hat{\delta}(x-x^\prime) + \Delta Z
\ee
The first term is the classical result $Z_{cl}= \left( \int d \phi  
U \right) |_{-\infty}^\infty$.
Substituting the mass renormalization $\mu^2_0 = \mu^2+(\delta \mu^2)_{susy}$ 
into $Z_{cl}$ leads to 
$\Delta Z = \frac{m}{ \lambda} (\delta \mu^2)_{susy}$.  
The second term is ultraviolet
divergent but this divergence will be removed by $\Delta Z$. 
The third term is the anomaly. It would vanish if we used 
$\delta(x-x^\prime)$ instead of $\hat{\delta}(x-x^\prime)$, and 
set $x=x^\prime$.
It is given by the integral 
\be
\label{int1}
\hbar
\int_{-L/2}^{L/2}dx \int_{-L/2}^{L/2} dx^\prime  U^{ \prime}[\phi(x)] 
i\int_{-\Lambda}^{\Lambda}
\frac{d^2p}{(2\pi)^2} \frac{i p_x}{p^2+m^2} 
\exp(ip_x(x-x^\prime)) \int_{-K}^{K}
\frac{dq}{2\pi} \exp(iq(x-x^\prime))
\ee
Note that we have used the propagators of the trivial sector. 
This is, of course, a crucial issue which we shall explain at the end of this section.
To compute this integral we use Maple and we take the 
$p_0,x,x^\prime$ integrals
first, then the $q^\prime=p+q$ integral, and finally the $p$ integral.
(For finite $L$, all $\Lambda$ and $K$ integrals are, of course, finite and unambiguous). 
In the limit of $L \to \infty$ the result reads
\be
\label{ps32}
\hbar \frac{U^{\prime}[\phi(\infty)]}{4 \pi} \left( 1+{\rm sign} (\Lambda-K ) \right)
\ee
In particular, for $\Lambda=K$ we find that $ {\rm sign} (\Lambda-K )$ in 
(\ref{ps32}) is equal to zero.

For the case 2), we obtain from the supercharge anticommutator 
\be
\label{fz}  
\langle Z_{ferm} \rangle = i \int_{-L/2}^{L/2} dx \int_{-L/2}^{L/2} 
dx^\prime 
 U^{ \prime} [\phi(x)]
\langle \bar{\psi}(x^\prime) \gamma^1 \psi(x) \rangle \hat{\delta}(x-x^\prime)
\ee
and the integral that we need to compute is exactly the same, 
as in the case 1). Hence, the fermion loop doubles the result of the boson loop, and the total result for the anomaly in the 
momentum cutoff scheme is
\be \hbar
\frac{U^{ \prime} (\infty)}{2 \pi} \left( 1+{\rm sign} (\Lambda-K ) \right)
\ee

The correct procedure is to use $K=\Lambda$ because in the theory
with $N$ modes we should use the delta function which corresponds
to exactly $N$ modes. This yields for the
anomaly   the following result
\be \hbar
\frac{U^{ \prime} (\infty)}{2 \pi}
\ee
Using the exact expression for $U(\phi)$, we recover the result (\ref{ura1}) 
for the central charge anomaly.
The BPS bound is saturated.

Note, however, that using under-regulated propagators ($\Lambda>K$),
would result in a result twice as large as ours. On the other hand,
under-regulating the delta function will give a vanishing result for the 
anomaly. 
This dependence of the anomaly on only ${\rm sign} (\Lambda-K )$ is a 
general feature. In the next section
we take closer look at the higher derivative regularization
scheme of ref \cite{shifman} and present three different 
supersymmetry currents (which differ from each other by an identically
conserved term)
each giving a different answer for the bosonic loops.  
We will establish a one-to-one correspondence with the
cases $\Lambda>K$, $\Lambda<K$ and $\Lambda=K$ discussed above, 
and argue 
that the higher derivative regularization scheme only properly regulates
one of these currents.

Let us now come back to the problem mentioned below (\ref{int1}), namely the
justification of using in (\ref{bz}) the propagators of the trivial sector.
One may view the propagator $\langle \eta^2(x) \rangle$ as a sum of terms of
the form of (\ref{f44}) [3,4]. The term with the trivial propagator 
corresponds to a loop graph with one insertion at $x$, whereas the 
other terms have one or more insertions of the background field 
combinations $V = - \frac{3}{2} \mu^2 \left( 
\tanh^2 \frac{\mu x}{\sqrt{2}} -1 \right) \eta^2 $. The term with
only the insertion at $x$ is canceled by the mass renormalization, because
we defined $\delta \mu^2$ precisely such that it canceled the tadpoles in the
trivial sector. The term with $\langle \eta(x) \partial_x \eta(x^\prime) \rangle $
in (\ref{bz}) without any $V$ insertions is the anomaly, and we computed it. 
We are left with 
all contributions in the term with $\langle \eta^2(x) \rangle$  
and in the term with $\langle \eta(x) \partial_x \eta(x^\prime) \rangle $ with one or 
more $V$ insertions. We claim that the contribution of their sum vanishes.
The crucial point is now to note that all these graphs are finite. We may therefore set
$\hat{\delta}(x-x^\prime)$ equal to ${\delta}(x-x^\prime)$, and obtain then a total derivative
\be
\label{prv1}
\frac{1}{2} \int_{-L/2}^{L/2} dx \partial_x \left[ 
U^\prime [\phi_K] \left(  \langle \eta^2(x) \rangle -  \langle \eta_0^2(x) \rangle \right)
\right]
\ee
This expression vanishes because for $ x \to \pm \infty$ the difference 
$\langle \eta^2(x) \rangle -  \langle \eta_0^2(x) \rangle $
vanishes. (Equivalently, all insertions vanish at large $x$). This proves
that it was correct to only use the propagators of the trivial sector for
the evaluation of (\ref{bz}).

\section{Comparison with previous results.}

In this section we comment on previous derivations of the one-loop corrections
to the mass and central charge of a bosonic or supersymmetric soliton in 1+1 
dimensions. 
For definiteness we discuss the kink but all results also hold for other models such
as the sine-Gordon model. 

\subsection{Derivative regularization.}

The Casimir approach of summing zero point energies is the 
oldest and most straightforward approach to obtaining 
$M^{(1)}$, but the result for naive momentum regularization
or mode number regularization depend on the boundary 
conditions for the quantum fluctuations \cite{pvn}.  
In ref. \cite{misha} it was proposed to consider instead 
the derivative of the two series in (\ref{lena}) w.r.t. $m$ because this reduces their linear divergence to a logarithmic 
divergence.
We call this scheme derivative regularization.
For the bosonic case one obtains then
\be
\frac{\partial}{\partial m} M^{(1)} = 
\frac{\partial}{\partial m} \frac{1}{2} \hbar \omega_B+
 \sum_n \frac{1}{2} \hbar 
\frac{m+k_n \frac{d k_n}{dm}}{(k_n^2+m^2)^{\frac{1}{2}}}
- \sum_n \frac{1}{2} \hbar 
\frac{m}{(k_n^2+m^2)^{\frac{1}{2}}}
+ \frac{\partial}{\partial m} \Delta M
\label{mtr}
\ee
Because the $k_n$ which solve
(\ref{h1}) depend on $m$, there is a $k_n \frac{d k_n}{dm}$ term. 
Since $\delta(k,m)$ depends only on 
$ k/{m}$, one has $k_n {d k_n}/{dm} = \frac{k_n^2}{mL} 
\frac{\partial}{\partial k} \delta \left( {k}/{m} \right)$.
There is one genuine bound state 
with $\omega_B=\frac{1}{2} \sqrt{3} m$.
This leads to
$$
\frac{\partial}{\partial m} M^{(1)} =  
\frac{\sqrt{3}}{4} \hbar m  + \frac{1}{2} \hbar 
\int_{-\Lambda}^{\Lambda } 
\frac{m+ \frac{1}{L} \frac{k^2}{m} \frac{\partial}{\partial k} \delta
\left( \frac{k}{m} \right) }{\sqrt{k^2+m^2}} \rho(k) dk - 
\frac{1}{2} \hbar 
\int_{-\Lambda}^{\Lambda } 
\frac{m}{\sqrt{k^2+m^2}} \rho_0(k) dk
$$
\be
+
\frac{\partial}{\partial m} \left( 
\frac{3m \hbar}{4 \pi} \int_{-\Lambda}^{\Lambda} \frac{dk}{\sqrt{k^2+m^2}}
\right)
\label{h2}
\ee
with $\rho(k)$ and $\rho_0(k)$ given in (\ref{mm1}).
From (\ref{h2})
we obtain $\frac{\partial}{\partial m} M^{(1)}$, and with the
renormalization condition $M^{(1)}(m=0)=0$, one obtains $M^{(1)}$ itself provided $M^{(1)}$ is linear in $m$ (see below).
The second and third term
combine into $\frac{1}{2m} \int_{-\Lambda}^\Lambda \frac{dk}{2 \pi} 
\sqrt{k^2+m^2} \frac{\partial}{\partial k } \delta \left( \frac{k}{m} \right)$.
They yield the same result as one would obtain by dividing the result for naive momentum cut-off by $m$, 
namely  $ \frac{1}{m} \left[ \frac{1}{2} \int_{-\Lambda}^\Lambda \frac{dk}{2 \pi} 
\sqrt{k^2+m^2} \frac{\partial}{\partial k } \delta \left( {k} \right) \right]$.
For the last term, however, taking the $\frac{\partial}{\partial m} $ 
derivative is not the same as dividing by $m$. The difference is due to
$\frac{\partial}{\partial m} $ acting on the $m^2$ in $(k^2+m^2)^{-1/2}$.
Without this extra term one obtains the result due
to naive momentum cut-off which does not agree with the standard semiclassical 
result [1] or, in the case of the sine-Gordon model, with the result obtained from the Yang-Baxter equation 
\cite{misha}. The extra term has the same effect
as inserting the cut-off functions $\Theta(\Lambda-k)$ or $f(k)$ in
the formula for the density of states, because
$ m \frac{3 m \hbar}{4 \pi} \int_{-\infty}^{\infty} \frac{-m}{ \left( k^2+m^2\right)^{3/2} } dk = 
- \lim_{k \to \infty} \frac{k \delta(k)}{2 \pi} = - \frac{3m}{2 \pi}$
according to (\ref{s17}). This result can also be written as $- \lim_{k\to \infty} k \delta (k)= -\lim_{k \to \infty}
-k \delta^{(1)}(k) = \frac{m^2}{\lambda} 
\frac{\partial}{\partial m} \delta \mu^2$ where $\delta^{(1)}(k)$ is the Born approximation .
 With this extra term one obtains the result in (\ref{ura1}). 

The basic approach of any mass or central charge computation is that one must 
regularize the theory from the very beginning, and only after that
perform {\it any} calculations. This was the basic strategy of the derivative regularization.
Namely, in \cite{misha} the system was first put in a box of length $L$, 
and the
$\frac{\partial}{\partial m}$ 
derivatives of the sums in (\ref{f3}) were taken. Afterwards, the
limit $L \to \infty$ was taken which converts sums to integrals.
This leaves the density of states $\rho(k)$
not differentiated with respect to $m$. This is crucial for the method of
\cite{misha}; differentiation of $\rho(k)$ would result to another
(incorrect) value of mass $M^{(1)}$.
 
Exactly the same result and conclusions are reached in the susy case, but to save space we will not 
explicitly demonstrate this here. 
Thus momentum regularization with a cut-off function is equivalent to derivative regularization in all cases. 

We now come back to our claim that $M^{(1)}$ should be linear in $m$. 
When the boundary conditions are such that divergencies remain,
say $M^{(1)}=A m + B m \ln \Lambda/m$ where $A$ and $B$ are constant, then the
$\frac{\partial}{\partial m}$ trick would lead to 
$\frac{\partial}{\partial m} M^{(1)}= A + B (\ln \Lambda/m -1)  $, and although the
renormalization condition $M^{(1)}(m=0)=0$ is satisfied, the answer is clearly
not given by $m \frac{\partial}{\partial m} M^{(1)}  (m=0) $. This is not 
an academic problem; for example, for the supersymmetric boundary conditions 
$\eta(L/2)=\eta(-L/2)=0$ and $\psi_1(L/2)=\psi_1(-L/2)=0$ all modes cancel pairwise, and thus $M^{(1)}=\Delta M$ in this case. 
Differential regularization
cannot be applied to this case without further modifications.
This clearly demonstrates that only those boundary conditions are allowed which 
produce a finite answer. The value of that finite answer can then be computed with differential regularization and yields
the correct correction to the mass.

\subsection{Mode regularization.}

Next we discuss mode regularization, i.e., the regularization scheme in which one
keeps equal numbers of modes in the nontrivial and trivial sector, both 
for bosons and for fermions.
To fix the idea we use a construction due to
Schonfeld \cite{schonfeld}, namely we consider a kink-antikink system 
$$
\phi_{K \bar{K}}(x) = \phi_K(x-\frac{L}{2}) 
\hspace{1cm} {\rm for}  \hspace{1cm} {0 \le x \le L}
$$
\be
\phi_{K \bar{K}}(x) = \phi_{\bar{K}} (x+\frac{L}{2})= - \phi_K(x+\frac{L}{2}) 
\hspace{1cm} {\rm for}  \hspace{1cm} {-L \le x \le 0}
\ee
\\
\epsfbox{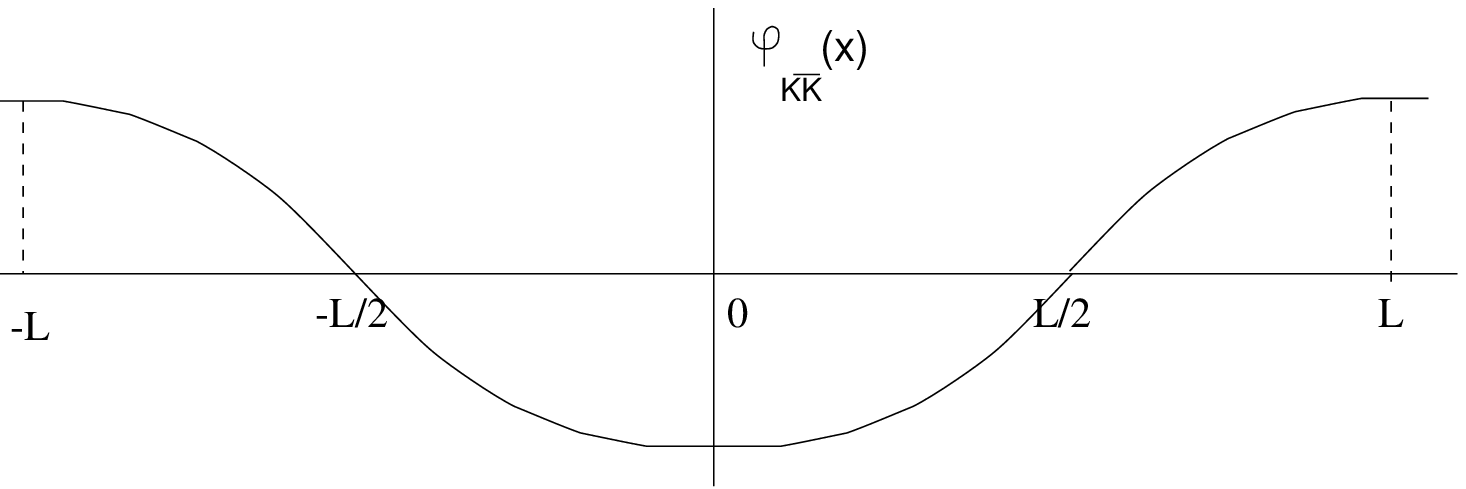}  \vspace{0.3cm}
\\
Figure 1. {\it The kink-antikink configuration.} \\ \\
This configuration clearly lies in the trivial sector, and for large
$L$ it becomes an exact solution. For finite $L$, the field equation
$-\partial^2_x \phi+UU^{ \prime}+ 
\frac{1}{2} U^{\prime \prime} \bar{\psi} \psi = 0$ is violated at
$x=0$ but the violation disappears in the limit $L \to \infty$ and we shall
disregard it. (All fields are massive).

One may begin with the trivial vacuum in a box of length $2L$,
and slowly turn on the configuration $\phi_{K \bar{K}} (x)$. The discrete
modes in the box move as one increases the background, and the 
difference of the distorted zero point energies
and the original zero point energies is the quantum correction to the work
done to create the $K \bar{K}$ system, namely $2 M^{(1)}$. 
This explains why and how mode regularization should work.  

Iterating the Dirac equation, one finds Schr\"odinger equations for
$\psi_1$ and $\psi_2$.
The potentials for the fluctuations $\eta$, 
$\psi_1$ and $\psi_2$ are all different 
\\ \\ \epsfbox{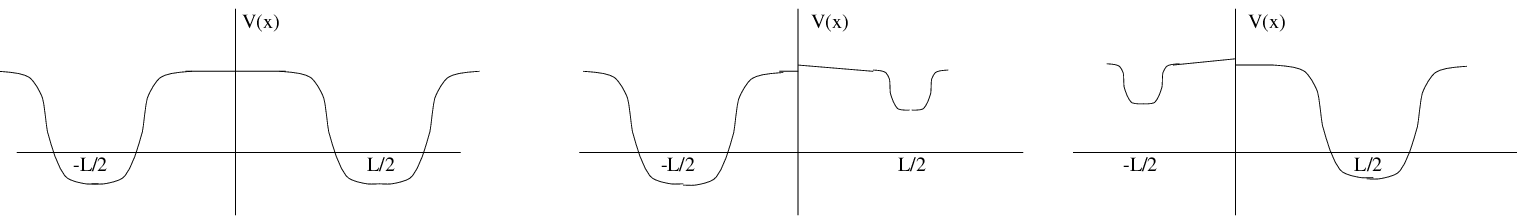} 
  
\hspace{1cm} $V(x)$ 
for $\eta$ \hspace{3cm} $V(x)$ for $\psi_2$ \hspace{3cm}
$V(x)$ for $\psi_1$ \\ \\
Figure 2. {\it Potentials for the bosonic and fermionic fluctuations in 
the kink-antikink system.} \\ \\
A plane wave incident from the left acquires a phase shift $\delta(k)$
in the deeper potential and a phase shift $\delta(k)+\theta(k)$ in the
shallower potential. Thus on the far right 
$\psi_1 \sim e^{i[kx+\delta(k)+\frac{1}{2} \theta (k)]}$, on the far left 
$\psi_1 \sim e^{i[kx-\delta(k)-\frac{1}{2} \theta (k)]}$ while near $x=0$
one has $\psi_1 \sim e^{i[kx-\frac{1}{2} \theta (k)]}$.
The phase shifts are given by        
$$ \epsfbox{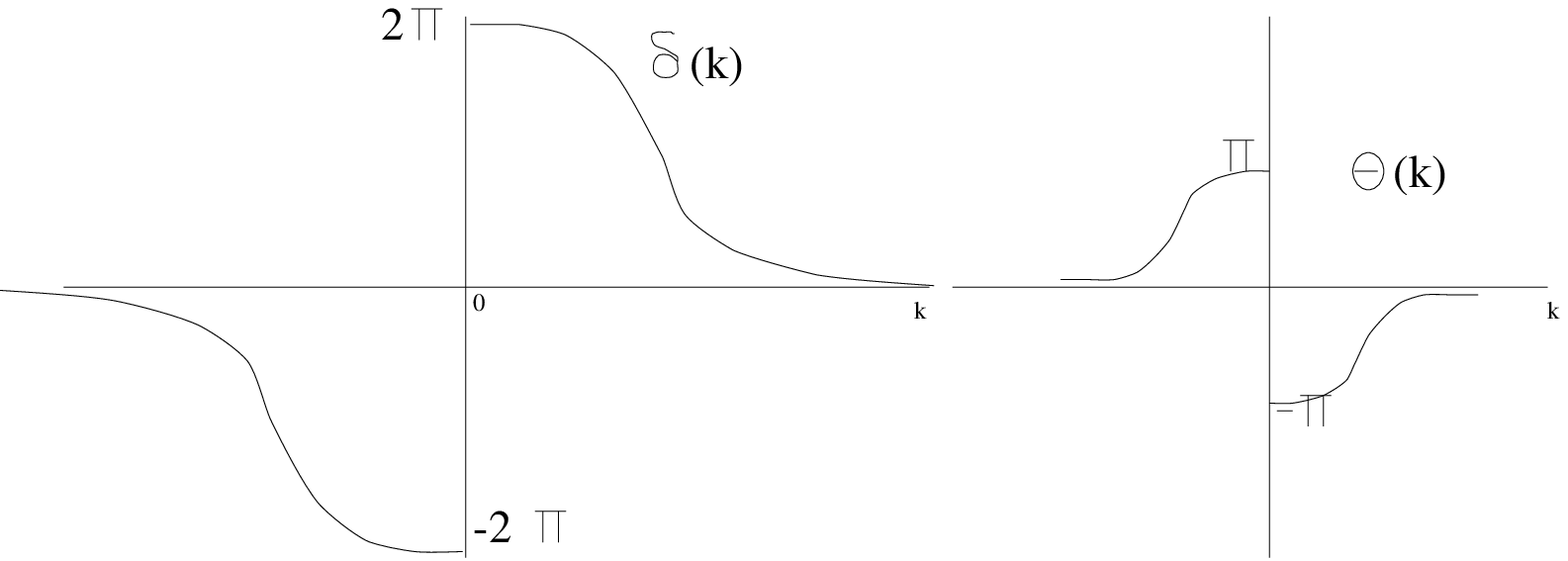} $$
$$
\begin{array}{c}
\delta (k) = 2 \pi \Theta(m^2-2k^2)- 2 \arctan \left( 
\frac{3 m k }{m^2-2 k^2}  \right) 
\hspace{0.3cm} {\rm for} \hspace{0.3cm} k>0 \\
\theta(k) = - 2 \arctan \frac{m}{k}  \hspace{0.3cm} {\rm for
} 
\hspace{0.3cm} k>0
\end{array} $$ 
\\ 
Figure 3. {\it The phase shifts for the bosonic and fermionic
fluctuations.} \\ \\
Imposing periodic boundary conditions on $\eta$ 
and $\psi_1$ in the $K \bar{K}$ system (to be discussed)  
leads then to the following quantization 
conditions on the momenta of the continuous spectrum
$$
2 k^B_nL+2 \delta(k^B_n) = 2 \pi n
$$
\be
2 k^F_nL+2 \delta(k^F_n) + \theta(k_n^F) = 2 \pi n
\ee
For $k_n^B$ the solutions start at $k^B_2=k^B_{-2}=0$, $k^B_{\pm 3}
= \pm \frac{ \pi}{ L} $, while 
for $k_n^F$ the solution starts at $k_{\pm 2}=\pm \frac{\pi}{2L}$,
$k_{\pm 3}= \pm \frac{3 \pi}{2 L}$.

In the discrete spectrum one might naively expect two bosonic zero modes, 
two fermionic zero modes, two bosonic bound states and two fermionic bound
states, because the kink and the antikink have each: one bosonic zero mode,
one bosonic bound state, one fermionic zero mode and one fermionic bound state.
However, this naive result for the kink-antikink system is incorrect. 
We can deduce the correct result by combining the information about the
bosonic $K \bar{K}$ system with the information about the susy $K \bar{K}$ 
system. In the bosonic  $K \bar{K}$ system we need 4 discrete modes in the nontrivial sector to compensate the loss of the 4 continuum modes with $n=\pm 1$ and $n=\pm 2$. In particular we need both zero modes of the  $K \bar{K}$ system,
even though one of them is unstable (it describes the attractive
force between a kink and antikink and corresponds to negative $\omega^2$). 
In the susy $K \bar{K}$ system, we need 3 discrete states in the 
fermionic sector to compensate the loss of the 3 continuum states with $n=0$
and $n=\pm 1$. In particular we need {\bf one} fermionic zero mode. The other fermionic zero mode 
does not correspond to a degree of freedom; rather, it
corresponds to a $Z_2$ gauge 
symmetry. We shall explain this elsewhere in more detail \cite{fred} where we also discuss the relation with 
sphalerons, but here we only note that the 2 states $(1+c_0)|0 \rangle$
and $(1+c_0) \tilde{c}_0 |0 \rangle$ are invariant under action 
with $c_0$ because $c_0^2=1$. The other 2 states $(1-c_0)|0 \rangle$
and $(1-c_0) \tilde{c}_0 |0 \rangle$ are also invariant under $c_0$, but
$\tilde{c}_0$ leads to transition between the first and fourth, or
second and third state. Thus the states $|0 \rangle$ and $c_0 |0\rangle$ are
``gauge related'', and $(1+c_0) | 0 \rangle $ is the gauge invariant state, and $c_0$ produces discrete gauge transformations. 
These results on the fermionic zero modes in the $K \bar{K}$ system correspond to results on the fermionic zero mode in the kink system obtained in \cite{shifman}, where 
it was stated that massive representations of the $N=(1,1)$ susy algebra 
which satisfy the BPS bound are {\bf one}-dimensional. 
Massive representations which do not satisfy the BPS bound are
two-dimensional. Hence, the result $M^{(1)}=Z^{(1)}$ can be explained in the
same way as Olive-Witten did for four-dimensional models.
In \cite{ritz}
the result was explained by reducing $N=(2,2)$ susy down to $N=(1,1)$. In \cite{fred}
we shall discuss these issues further.  

Thus, in the kink-antikink system there is one discrete mode less
in the fermionic sector than in the bosonic sector, and therefore mode regularization (meaning keeping equal number of bosonic and fermionic modes) requires
that one should include one extra mode of the continuum spectrum of the fermions. This is
exactly what Schonfeld did in 1979 \cite{schonfeld}. His motivation was not to use mode regularization, 
but rather he noticed that for large momenta $k_n$
the corresponding bosonic and fermionic levels lie next to each
other, though the fermionic level is always a little bit higher than 
the bosonic one. Schonfeld picked all levels up to a certain momentum,
which is the idea of momentum cut-off. From the point of view of 
mode cut-off this implied that he took one more mode for the fermions.
There is still a problem left, not discussed by him: there is a
(very small) probability that the momentum cut-off falls
precisely between nearly degenerate boson and fermion levels. Continuing 
this line of thought, one would then be led to wonder whether a kind
of averaged momentum cut-off would include a contribution of the form ``small
level difference times large extra mode''. We shall not pursue this issue and return to mode regularization.

We begin by stressing that in the susy case we always choose the same boundary conditions
in the sector without $K \bar{K}$ configuration for the bosons and fermions, 
in order that the free bosonic and fermionic zero modes cancel. We need then only to 
consider modes in the $K \bar{K}$ system.

The mass of the susy kink-antikink system with periodic boundary conditions on $\eta$ and $\psi_1$ is then given by
$$
M_{K \bar{K}} \approx 2 M_K = \left[ 2 \frac{1}{2} \hbar \omega_B +2 \times  0 +
\frac{1}{2} \hbar m + \sum_{n=3}^N  \hbar \sqrt{(k^B_n)^2+m^2} \right]
$$
$$
- \left[  2 \frac{1}{2} \hbar \omega_B + 0 + \sum_{n=2}^N \hbar
\sqrt{(k^F_n)^2+m^2 }  \right] + 2 \Delta M_{susy}
$$
$$ 
=
\frac{1}{2} \hbar m - \hbar m + \hbar \sum_{n=3}^N \left( \sqrt{(k^B_n)^2+m^2}-\sqrt{(k^F_n)^2+m^2 } \right) + 2 \Delta M_{susy} 
$$
$$
= -\frac{1}{2} \hbar m + \hbar \int_0^\Lambda \frac{dk}{\pi} 
\left(  
\frac{d}{dk} \sqrt{k^2+m^2} 
\right)
\left(
\frac{1}{2} \theta (k)
\right) + 2 \Delta M_{susy}
$$
\be
= -\frac{1}{2} \hbar m + \frac{\hbar}{2 \pi} \sqrt{k^2+m^2} \theta(k) |_0^\infty 
= - \frac{1}{2} \hbar m - \frac{\hbar m}{\pi} + \frac{1}{2} \hbar m = 
 - \frac{\hbar m}{\pi}
\ee
We used that $k_n^B-k_n^F = \frac{1}{2L} \theta(k_n)$ and
\be
-\int_0^\Lambda \frac{dk}{4 \pi} \sqrt{k^2+m^2} \frac{d}{dk} \theta (k) + 
\Delta M_{susy} = 0
\ee
a result which was already obtained in eq. (59) of \cite{pvn}. The result for the mass
of the susy kink is $M_K=-\hbar 
\frac{m}{2 \pi}$, which is in agreement with (\ref{ura1}).

If one takes periodic boundary conditions for $\psi_2$ instead of $\psi_1$,
the result is clearly the same. Choosing anti-periodic boundary conditions 
for $\eta$ and $\psi_1$ (or $\psi_2$) in the $K \bar{K}$ 
sector one also finds the 
correct result. 
One can even use periodic boundary conditions for $\eta$ and 
anti-periodic boundary conditions for $\psi_1$ (or $\psi_2$), or vice-versa,
they all give the same result.
This agrees with one's intuition that the result
should not depend on the boundary conditions in the trivial sector.

One can repeat the same calculations for the bosonic kink. There are now
4 discrete states in the $K \bar{K}$ sector. One should always use the same boundary conditions in the $K \bar{K}$
system as for the free bosons, but it does not matter what these boundary conditions are. If one uses different
boundary conditions in the $K \bar{K}$ system and in the free system, 
one finds an extra positive contribution which
can be interpreted as the energy needed to twist fields at the boundaries. (For example, anti-periodic conditions
in the $K \bar{K}$ system and periodic in the free system yield an extra energy $\frac{\hbar m}{2}$ for $M^{(1)}$).
For definiteness we consider the case of periodic bosonic boundary conditions in the $K \bar{K}$ system and the free 
system. Then one has
$$
2 M^{(1)} = \left( 2 \frac{\hbar}{2} \omega_B + 2 \times  0 + \frac{1}{2} \hbar m +
\sum_{n=3}^N \sqrt{(k_n^{B})^2+m^2} \right) - \sum_{n=-N}^N  \hbar
\sqrt{(k_n^0)^2+m^2}+ 2 \Delta M_{bos}
$$
$$
= \hbar \omega_B - 2 \hbar m + \hbar \sum_{n=3}^N \left( \sqrt{(k_n^{B})^2+m^2} - \sqrt{(k_n^0)^2+m^2} \right) + 2 \Delta M_{bos}
$$
$$
=\hbar \omega_B - 2 \hbar m + \hbar \int_0^\Lambda  \frac{dk}{\pi} \frac{d \omega}{dk} (-\delta(k)) + 2 \Delta M_{bos}
$$ 
\be
= \hbar \omega_B - 2 \hbar m - \hbar \frac{\omega \delta(k)}{\pi} |_0^\Lambda + \left[
\int_0^\Lambda \frac{dk}{\pi} \omega \delta^\prime (k) + 2 \Delta M_{bos} 
\right]
= \hbar \omega_B - \frac{3 m \hbar}{\pi} 
+ 2 \left[ - \frac{\sqrt{3}m \hbar}{6} \right]
\label{s99}
\ee
with $\omega_B = \frac{\sqrt{3}}{2}m$ we get indeed (\ref{ura1}). 
We used $\hbar \int_0^\Lambda \frac{dk}{2 \pi} \omega \delta^{\prime} (k) + \Delta M_{bos}= - \frac{\hbar \sqrt{3} m}{6}$.

We now read off from the result for the $K \bar{K}$ system which boundary conditions for the $K$ system yield the correct result.
Periodicity of $\eta$ on the double interval allows both periodicity
and anti-periodicity on the single interval 
($\eta(0)=\pm \eta(L) = \pm \eta (-L)$). The quantization conditions for
$\eta$ around a kink is then 
\be
\label{lk3}
k_n^BL+\delta(k) = (2 \pi n  \hspace{0.3cm}{\rm or} 
\hspace{0.3cm} 2 \pi n + \pi)= \pi m
\hspace{1cm} (m \ge 0  \hspace{0.3cm}{\rm and} 
\hspace{0.3cm} m < 0)
\ee
For $\psi_1$ periodicity on the double interval led to 
\be
\label{mm2}
k_n^FL +\delta(k_n^F)+\frac{1}{2} \theta(k_n^F)=\pi n
\ee
and
this quantization condition can also be
obtained for a single kink by imposing  both twisted boundary conditions of ref \cite{misha}
\be
\psi_1(0)=\psi_2(L) \hspace{0.4cm} {\rm and} \hspace{0.4cm} 
\psi_2(0) = \psi_1(L) \hspace{0.4cm} : \hspace{0.4cm} kL+\delta(k)+
\frac{1}{2} \theta(k) = 2 n \pi
\label{lk1}
\ee
as well as
\be
\label{lk2}
\psi_1(0)= -\psi_2(L) \hspace{0.4cm} {\rm and} \hspace{0.4cm} 
\psi_2(0) = -\psi_1(L) \hspace{0.4cm} : \hspace{0.4cm} kL+\delta(k)+
\frac{1}{2} \theta(k) = 2 n \pi+\pi
\ee
These boundary conditions correspond to the two 
$Z_2$ symmetries of the action $\phi \to - \phi$ and either 
$\psi \to \tau_3 \psi$
or $\psi \to - \tau_3 \psi$ \cite{misha}.
We have thus given an alternative explanation of the twisted boundary conditions, which were derived in \cite{misha} requiring absence of a genuine boundary
in a box. 
Combining (\ref{lk1}) and (\ref{lk2}) yields (\ref{mm2}).
Thus one obtains in the kink 
sector the correct
mass if one takes the average of  both periodic 
and anti-periodic boundary conditions for $\eta$,
and takes the average of the two sets of twisted boundary conditions on the fermions.

Another set of 
correct boundary conditions on the double interval are the supersymmetric 
conditions of 
\cite{schonfeld} with $\eta(-L)=\eta(L)=0$ and $\psi_2(-L)=\psi_2(L)=0$
(or $\psi_1(-L)=\psi_1(L)=0$ ). For
the kink on an interval of length $L$ this implies for the bosons both  
Dirichlet and
Neumann boundary conditions: $\eta(L)=0$ and further both $\eta(0)=0$ and 
$\frac{\partial}{\partial x}\eta(0)=0$. 
This leads to $kL+\delta(k) = n \pi$ ($k>0$) and  
$kL+\delta(k) = m \pi+\pi/2$ ($k>0$) which can be combined into 
$kL +\delta(k) = n \pi /2 $ for $n>0$. The density of the spectrum
is the same as in (\ref{lk3}).  
For the fermions on the double interval the wave function $\psi_2$ is given by either
$\psi_2 \sim \sin[kx \pm \delta(k) \pm 
\frac{1}{2} \theta(k)]$ or
$\psi_2 \sim \cos[kx \pm \delta(k) \pm \frac{1}{2} \theta(k)]$ 
at $x=\pm L$ and the quantization conditions become
\be
\left.
\begin{array}{c}
2 k L + 2 \delta(k) + \theta(k) = 2 \pi n \\
2 k L + 2 \delta(k) + \theta(k) = (2n+1) \pi \\
\end{array}
\right\}  
\hspace{0.5cm} kL+\delta(k) +\frac{1}{2} \theta(k) = \frac{n \pi}{2} 
\ee
Again we can deduce from these boundary conditions for $\psi_2$ on the double 
interval a correct set of boundary conditions for $\psi_1$ and $\psi_2$
in a kink background on the interval between $x=0$ and $x=L$.
We find then that both $\psi_1$ and 
$\psi_2$ vanish at $x=L$ but satisfy both Dirichlet and Neumann conditions at $x=0$
\be
\begin{array}{c}
\psi_1(L)=0, \hspace{1cm} \psi_1(0) =0 , \hspace{1cm} {\rm and } \hspace{1cm} 
\psi_1^\prime(0) = 0 \\
\psi_2(L)=0, \hspace{1cm} \psi_2(0) =0 , \hspace{1cm} {\rm and}  \hspace{1cm} 
\psi_2^\prime(0) = 0 \\
\end{array}
\ee
Because the mass of the kink is half of that of the kink-antikink system,
we must average over this conditions
\be
M^{(1)} = \sum \frac{1}{2} \hbar \omega^B_n -
\left( \sum \frac{1}{4} \hbar \omega^{\psi_1}_n +
\sum \frac{1}{4} \hbar \omega^{\psi_2}_n
\right) + \Delta M
\ee
This yields indeed the correct result.
Note that susy is completely broken in the $K \bar{K}$ system: one half of susy is broken by the kink and the other half by the antikink.
This explains why the bosonic and fermionic frequencies do not cancel each other and it also explains why the conditions 
 $\psi_2(-L)=\psi_2(L)=0$ are equivalent to $\psi_1(-L)=\psi_1(L)=0$.

\subsection{The Born approximation approach.}
In ref \cite{graham} the problem of computing the one loop corrections
to the bosonic and susy kink was approached by rewriting the sums
$ \sum \frac{1}{2} \hbar \omega_n - \sum \frac{1}{2} \hbar \omega_n^0 $ in 
terms of the effective action, with 
one-loop graphs with flat-space propagators 
and vertices $- \frac{3}{2} \mu^2 \left( 
\tanh^2 \frac{\mu x}{\sqrt{2}} -1 \right) \eta^2 $ 
for the kink background 
\be
\label{f44}
\sum \frac{1}{2} \hbar \omega_n - \sum \frac{1}{2} \hbar \omega_n^0 =
\hspace{7cm}
\ee
\nopagebreak
\setlength{\unitlength}{0.16mm}
\begin{picture}(0,0)(-510,-55)
\put(1,8){\shortstack{(}}
\put(40,15){\circle{28}}
\put(80,8){\shortstack{+}}
\put(120,15){\circle{28}}
\put(120,3){\circle*{10}}
\put(160,8){\shortstack{+}}
\put(200,15){\circle{28}}
\put(200,3){\circle*{10}}
\put(200,27){\circle*{10}}
\put(240,8){\shortstack{.}}
\put(250,8){\shortstack{.}}
\put(260,8){\shortstack{.}}
\put(270,8){\shortstack{)}}
\put(292,8){\shortstack{---}}
\put(340,15){\circle{28}}
\end{picture}
The vertices are fixed functions of $x$, but one can consider vertices which
depend on a function $J$ by adding a source term $\int J \eta dx$ to the action \cite{graham}. This
term does not appear in the effective action, but now the kink solution
$\phi_K(x,J)$ depends on $J$, and so therefore do the vertices.
At this point no renormalization has yet been introduced.

Next the authors of \cite{graham} used the relation
$\rho(k)-\rho_0(k) = \frac{1}{\pi} \frac{\partial}{\partial k} \delta(k)$
for the continuous spectrum
to find another expression for (\ref{f44})
\be
\sum \frac{1}{2} \hbar \omega_n - \sum \frac{1}{2} \hbar \omega_n^0 =
\frac{1}{2} \hbar \sum_{ds} \omega_j+ \int_0^\infty \frac{dk}{\pi} \frac{1}{2} \hbar  \sqrt{k^2+m^2}  
\frac{\partial}{\partial k} \delta(k)
-\frac{\hbar m}{4}
\label{f45}
\ee
Because one subtracts $\rho(k)-\rho_0(k)$ {\bf at the same $k$}, one is 
implicitly using momentum cut-off.  
The sum over discrete states $\omega_j$ is a 
sum over the zero point energies of the usual bound states
(including the zero mode for translation) plus a contribution from a 
``half bound state'' $\frac{\hbar m}{4}$ 
at the edge of the continuous spectrum, 
with $k=0$ but with half the usual energy, $\omega = \frac{1}{2} m$,  
which is 
removed in (\ref{f45}) by the last term $-\frac{ 1}{4} \hbar m$. 
One adds and subtracts this half-bound state in order to be able
to apply later Levinson's theorem.
The idea is now to make a Born expansion of $\delta(k)$ , namely 
$\delta(k) = \delta^{(1)}(k) + \delta^{(2)}(k) + ...$
where $ \delta^{(1)}(k) $ is the usual Born approximation of 
the phase shift in potential $V(x)$
\be
\delta^{(1)}(k) = - \frac{\hbar}{k} \int V(x) dx
\ee
and to identify the terms with $l$ vertices in (\ref{f44})  with 
the terms containing $\delta^{(l)} \left( {k}/{m} \right) $ in (\ref{f45}).  
However, this identification is clearly inconsistent as the contribution of 
$\delta^{1}(k)$ to (\ref{f45}) is infrared divergent. 
To remove the infrared divergence, use was made of
Levinson's theorem for the relation between the number of bound states 
$n_+$ and $n_-$ in the channels with positive and negative parity 
and the phase shift at $k=0$ 
\be
\delta^+(k=0) = \pi (n^+-\frac{1}{2} ) ; \delta^-(k=0) = \pi n^- 
\ee
For the bosonic kink, the collective mode with $\omega=0$ and the half-bound state
with $\omega = \frac{1}{2} m$ are included in $n^+$, so
$n^+=1+\frac{1}{2}$, and $n^-=1$ contains the true bound state.
Levinson's theorem can be rewritten as
\be
\sum_{ds} 1 + \int_0^\infty \frac{dk}{\pi} \frac{\partial}{\partial k}  
\left[
\delta^+ (k) + \delta^- (k)
\right] - \frac{1}{2} =0
\label{f46}
\ee
Multiplying (\ref{f46}) by $m$ and subtracting from (\ref{f45}) leads to 
\be
\sum \frac{1}{2} \hbar \omega_n - \sum \frac{1}{2} \hbar \omega_n^0 =
\frac{1}{2} \hbar \sum_{ds} (\omega_j-m) + \int_0^\infty \frac{dk}{2 \pi} 
( \omega -m) \frac{\partial}{\partial k} \left[
\delta^+ (k) + \delta^- (k)
\right]
\label{f47}
\ee
Now the Born approximation does not contain an infrared divergence because 
the factor $\omega - m$ is proportional to $k^2$ near $k=0$.
For this series in (\ref{f47}) it is claimed that the term in (\ref{f44}) with $l$ vertices corresponds to the
term in (\ref{f47}) with the $l^{\rm th}$ Born approximation. 

At this point the problem of regularization and renormalization is 
addressed in [3,5].
In particular, the term in (\ref{f47}) with 
$\delta^{(1)}(k) $ should exactly 
correspond to the graph with one vertex, and by
subtracting
this term, one should have performed mass renormalization. The two terms in
(\ref{f44}) without any vertex should cancel without leaving 
any finite part.

There remains the sum over all $\delta^{(l)}$ with $l \ge 2 $, or equivalently,
the sum over all graphs with $l \ge 2$ vertices. This sum is finite and can be 
evaluated (by using the expressions for $\delta(k)$ and $\delta^{(1)}(k) $.)
The result agrees with (\ref{ura1}). 

The proposed identification in [3,4,5] between individual graphs in the 
effective action and individual terms in the Born expansion of the
phase shift formula modified by Levinson's theorem is, at least for
the finite graphs with $l \ge 2$, plausible but not proven.
(We know of no reference, only of lore).
In particular, it is not specified in [3,4,5] which regularization 
scheme is supposed to be
used such that this identification becomes true.
In fact, we claim that for $l=0$ and $l=1$ the identifications do not
hold in general.
The contribution 
from the first Born approximation is given by (see equation
(12) in \cite{graham})
\be
\hbar
\int_0^\infty \frac{dk}{2 \pi} [\omega(k)-m] \frac{d}{dk} \frac{ \langle V
\rangle }{k}
\ee
On the other hand, the contribution from the graph with one insertion
is the counterterm and given by
\be
3 \hbar m \int_0^\infty \frac{dk}{2 \pi} \frac{1}{\sqrt{k^2+m^2}}
\ee
To make sense of those two divergent expressions one
must choose a regularization scheme. If one chooses a momentum cut-off,
these two contributions differ by a boundary term (partially
integrating one, one gets the other plus a boundary term).
As shown in \cite{pvn}, the results for the mass then
differ. On the other hand, for regularization schemes where these boundary terms vanish, the identification clearly holds.

Recently it has been checked that with dimensional regularization
the result for the term with $\delta^{(1)}$ is indeed the same as the 
result for the graph with one vertex \cite{recent}. 
However, as we already stated, 
this equality is scheme dependent: for example, it
does not hold for momentum cut-off. Presumably, for momentum cut-off one would
need to subtract not only the term with $\delta^{(1)}$ but also a 
finite amount; the final counter term would then be equal to the counter
term with $\delta \mu^2$ of the Casimir approach with momentum cut-off.
Apparently dimensional regularization is particularly well suited 
for the phase-shift method because it corresponds to minimal subtraction 
with only $\delta^{(1)}$.

\subsection{Higher derivative regularization for the computation of the central charge.}

In \cite{shifman} a higher-derivative supersymmetry-preserving regularization scheme was proposed to compute $Z^{(1)}$. It is the counterpart of the derivative regularization scheme in \cite{misha} which preserves the $Z_2$ symmetry.
As we have already discussed, 
in the susy case there are extra contributions to the
energy at the boundaries which one must subtract to obtain the kink mass,
once this is understood, both schemes are equally correct. Actually, we
claim that higher derivative regularization does not regulate.
The authors of \cite{shifman} have told us that they are aware of this
problem, and that by imposing in addition
current conservation the regularization is restored. In any case, the problem
is interesting, and we discus it now.

The action on which the higher derivative scheme is based can
be written in terms of superfields, so that the supersymmetry is manifestly
preserved by this approach. 
The Lagrangian density in components is given by 
\begin{equation}
L=\frac{1}{2} \left\{ - \partial_\mu \phi  \left( 1 - \frac{\partial^2_x}{M^2} 
\right) \partial^\mu
\phi - \bar{\psi}  \left( 1 - \frac{\partial^2_x}{M^2} \right) 
  {\not \partial} \psi 
+ F  
 \left( 1 - \frac{\partial^2_z}{M^2} \right) F + 2 U(\phi) F - U^{\prime} (\phi) 
\bar{\psi} \psi \right\} 
\end{equation}
The terms with $\partial^2_x/M^2$ are separately supersymmetric,
and since there are no higher time derivatives, the scheme is still canonical.
The higher derivative terms result in extra powers of
momentum in the denominator of the momentum space propagator, thus
rendering the loops with such propagators and with regular vertices finite.
However, we must compute loops with currents at vertices.
Such vertices contain extra powers of
momenta which destroy the effect of regularization and make the
integrals in general divergent and ill-defined. This is exactly the reason
why the higher derivative regularization scheme does not work 
for gauge theories at the one loop level. (The improvement in convergence due
to the propagators is undone by the decrease in convergence due to 
vertices, and in one-loop graphs there are as many propagators and vertices.
The method does  work at
higher loop). 

The supersymmetry current that one computes
with the Noether method is not unique. 
One may add extra terms to the susy current which themselves are
identically conserved. In particular one can find a current among 
this class of currents such that the one loop diagrams with this improved 
current are properly regulated by the higher derivative regularization
scheme.
We call this current $j^\mu_{imp}$; it is different from the current
$j^\mu_{SVV}$ used in ref \cite{shifman} and from the current $j^\mu_{N}$
which we computed with the Noether method generalized to higher
derivative schemes (we will discuss this
method somewhere else \cite{me}). 

Consider the following three conserved currents:
\begin{equation}
\label{jn}
j_{N}^\mu = - (\partial_\nu \phi) \gamma^\nu \gamma^\mu  \left( 1 - \frac{\partial^2_x}{M^2} \right) \psi -U \gamma^\mu \psi 
-
\frac{\delta^\mu_x}{M^2} \left[ ( \Box \phi ) 
\stackrel{\leftrightarrow}{\partial_x}  \psi + 
 F {\stackrel{\leftrightarrow}{\partial_x}  } (\not \partial \psi) \right]
\ee
\be
\label{jsvv}
j_{SVV}^{\mu} = j_{N}^\mu + \frac{1}{M^2} \epsilon^{\mu \nu } \partial_\nu
 (\not \partial \phi  \stackrel{\leftrightarrow}{\partial_x} \gamma^0 \psi)
\ee
\be
\label{jimp}
j_{imp}^{\mu}=j_{N}^{\mu} -\frac{1}{M^2} \epsilon^{\mu \gamma}
 \partial_\gamma 
\left[ (\partial_\nu\phi) \gamma^\nu \gamma^0 (\partial_x \psi) \right]
\ee
For the charges we need the components with $\mu=0$; these read
\begin{equation}
j_{N}^0 = - ( \partial_\nu \phi) \gamma^\nu \gamma^0  \left( 1 - \frac{\partial^2_x}{M^2} \right) \psi - U \gamma^0 \psi
\end{equation}
\begin{equation}
j_{SVV}^{0} = - \left( \partial_\nu   \left( 1 - \frac{\partial^2_x}{M^2} 
\right) 
\phi \right) \gamma^\nu \gamma^0  \psi -U \gamma^0 \psi
\end{equation}
\be
\label{jimp00}
j_{imp}^0 = -( \partial_\nu \phi) \gamma^\nu \gamma^0  \psi -U \gamma^0 \psi -\frac{1}{M^2} ( \partial_\nu \partial_x \phi) \gamma^\nu \gamma^0  \partial_x \psi
\ee
Consider now the supercharge algebra.
Using for $j^{0}_{imp}$ 
standard canonical commutation relations for the higher derivative 
scheme (note that there are no higher derivatives w.r.t. time, hence 
canonical methods are as usual) we obtain a result for the central charge $Z$
which looks like (\ref{bz}) and (\ref{fz}) but with a different meaning
of the regularized delta function. The bosonic contribution is given by
\be
\label{me11}
 \int_{-L/2}^{L/2} dx \int_{-L/2}^{L/2} dx^\prime 
U^{ \prime} [\phi(x)]
\langle \eta(x) \partial_{x^\prime} \eta(x^\prime) \rangle \hat{\delta}(x-x^\prime)
\ee
where 
\be
\label{deltahat}
\hat{\delta}(x) = \frac{1}{1-\frac{\partial_x^2}{M^2}}\delta_{Dirac}(x)= \frac{M}{2}
\exp(-M|x|)
\ee
The expression in (\ref{me11}) is finite, but it does not vanish due to symmetric integration because of the factor 
$e^{ik(x-x^\prime)} e^{-M|x-x^\prime|}$. The contribution from the last two terms in (\ref{jimp00}) vanishes for $M \to \infty$ because 
it contains a conditionally convergent integral $\int \frac{d^2k k_\nu k_x}{k^4}e^{ik(x-x^\prime)}$ and a factor 
$\partial_x e^{-M|x-x^\prime|}$. 
Note that both the commutator $[\dot{\phi}(x),\phi(x^\prime)]$ and 
$\{ \bar{\psi}(x), \psi(x^\prime) \}$ give a $\hat{\delta}(x-x^\prime) $. 
Note also that both the propagators and the delta functions are regulated. 

As in section 3, the fermionic contribution just doubles the bosonic contribution 
(\ref{me11}).
An easy 
computation gives the same correct value for the total anomaly  
\be
\label{sv2r} \hbar
\frac{U^{\prime} [\phi(\infty)]}{2 \pi} 
\ee

Consider now what one gets from the
current $j^0_{SVV}$ in \cite{shifman}.
The bosonic contribution to the central charge has now an extra term 
proportional to $\partial^2_{x^{\prime}}/M^2$. The bosonic contribution to 
$Z^{(1)}$ is
\be
\label{sv1r}
 \int_{-L/2}^{L/2} dx \int_{-L/2}^{L/2} dx^\prime 
U^{\prime} [\phi(x)]
\langle \eta(x) \left( 1-\frac{\partial^2_{x^\prime}}{M^2} \right) \partial_{x^\prime} \eta(x^\prime) \rangle \hat{\delta}(x-x^\prime)
\ee
The derivatives $\partial^2_{x^\prime}$ in the extra term undo the regularization provided by the propagator. 
One is left with a divergent integral  $\sim \int \frac{k e^{ik(x-x^\prime)} d^2 k}{k^2}$. We recognize here 
the failure of higher-derivative regularization at the one-loop level,
well known from gauge theories. To remedy the situation, one needs an additional regularization for the one-loop graphs, for example Pauli-Villars regularization or imposing current conservation. The authors of \cite{shifman} were well aware of this situation \cite{pismo}. Strictly speaking, their derivation of $Z^{(1)}$ is ambiguous because it is based 
on (\ref{jsvv}), whereas our derivation based on (\ref{jimp}) happens to be unambiguous because all
integrals are conditionally convergent. It makes a big difference where the extra derivatives act: one can suffer 
just one extra derivative on $\hat{\delta}$ and another in the $\phi$ loop, and this is what (\ref{jimp}) accomplishes. 

Although the momentum integral of the propagator in (\ref{sv1r})
is not regulated one can still evaluate this expression by first integrating over $x$, $x^\prime$ and then over $p$. 
One gets then a finite answer which is the correct result.
The fermionic contribution due to the anomaly vanishes for this current
because one obtains an exact Dirac delta function in this case. 

Let us now consider the Noether current $j_{N}$. Then
we find for the bosonic contribution to $Z^{(1)}$ 
\be
\int_{-L/2}^{L/2} dx \int_{-L/2}^{L/2} dx^\prime 
U^{\prime} [\phi(x)]
\langle \eta(x) \partial_{x^\prime} \eta(x^\prime) \rangle {\delta}(x-x^\prime)
\ee
Note that we now obtain an ordinary Dirac delta function instead of 
$\hat{\delta}(x-x^\prime) $. Setting $x^\prime$ equal to $x$, the result for
$\langle \eta(x) \partial_x \eta(x) \rangle$ vanishes and we have a zero 
value for the bosonic contribution to $Z^{(1)}$. 
However, the delta function $\delta(x-x^\prime)$ can be regulated in different ways giving different
answer (for example as in (\ref{deltahat}). Thus we consider the computation based on $j^0_{N}$ as ambiguous.
The fermionic 
contribution is divergent, but evaluating as in (\ref{sv1r}), one finds the right answer.

All the currents give the right answer (if, at least, one performs the evaluation as we have outlined), 
but the contributions to the value
of an anomaly come from different sectors for different currents. 
The improved current $j^0_{imp}$ yields $\frac{1}{2}$ of its contribution in the bosonic sector and  $\frac{1}{2}$
in the fermionic sector. The current $j^0_{SVV}$ used in \cite{shifman} yields only contributions in the bosonic sector while
the current $j^0_N$ yields only contribution in the fermionic sector. 

This 
may be understood along the lines of section 3. There we have noticed that only
three different answers may be computed for a given integral with anomaly,
depending on whether the integral is properly regulated or not. The 
only current here which is properly regulated is $j^\mu_{imp}$. 

It was pointed out \cite{pismo} that the topological currents that we 
present here are not local. ( The $\hat{\delta}(x-x^\prime)$ are nonlocal
objects). One of us will argue elsewhere \cite{me} that
replacing the nonlocal $\hat{\delta}(x-x^\prime)$ by local  ${\delta}(x-x^\prime)$ leads to inconsistencies.

Finally, we discuss an important issue, namely that on general grounds $M^{(1)}=Z^{(1)}$,
so that no separate calculation of $M^{(1)}$ would be needed if $Z^{(1)}$ is known. 
The argument is based on properties of representations of 
the susy algebra $Q_+^2=H+P$,
$Q_-^2 = H-P$ and $\{ Q_+,Q_- \}=Z$, which can be rewritten in terms of $Q_1=Q_++Q_-$ and $Q_2=Q_+-Q_-$ as 
$Q^2_1=H+Z$, $Q^2_2=H-Z$ and $\{ Q_1,Q_2 \}= 2 P =0$ 
for $\langle P \rangle =0$. 
(Although for finite $L$ the charges $Z$ and $H$ do not commute with each other , these are boundary effects,
and in the infinite volume $Z$ commutes with all charges).   
If $H \neq |Z|$ then $Q_1 |kink \rangle \neq 0$
and  $Q_2 |kink \rangle \neq 0$, and one might think that this 
leads to a susy multiplet of four instead of two states. 

The situation with BPS and non-BPS representations in 1+1 dimensions has been
confusing for some time. Consider massive non-BPS representations.
Even though $Q_1 |kink \rangle $ and $Q_2 | kink \rangle $ are nonzero,
it could be that they are proportional to each other\footnote{
In ref. \cite{shifman} 
it is stated that ``... we do not have multiplet shortening in two
dimensions''. However, in a footnote
added in print it is stated that ``
Actually, the supermultiplet is reducible: the states $(|sol_b \rangle \pm |sol_f \rangle )/ \sqrt{2}$
present two one-dimensional representations of the superalgebra''. This implies in our opinion that multiplet shortening does occur,
after all. 
We demonstrate this explicitly in the text.
}. 
Consider for example the following $2 \times 2$ matrix representation 
of $Q_1$ and $Q_2$
\begin{equation}
Q_1=\left( \begin{array}{cc} 0 & \sqrt{H+Z} \\ \sqrt{H+Z} & 0 \\ \end{array} \right), \hspace{2cm}
Q_2=\left( \begin{array}{cc} 0 & i \sqrt{H-Z} \\ -i \sqrt{H-Z} & 0 \\ \end{array} \right)
\end{equation}
As required, $Q_1$ and $Q_2$ anticommute.
If $H>|Z|$, the state $\left( {\tiny \begin{array}{cc} 1 \\ 0 \\ \end{array} } \right) $ is annihilated by 
$i \epsilon Q_1 + Q_2 $ where $\epsilon = [(H-Z)/(H+Z)]^{1/2}$. Thus, although 
$Q_1 \left( {\tiny \begin{array}{cc} 1 \\ 0 \\ \end{array} } \right) 
\ne 0$ and  
$Q_2 \left( {\tiny \begin{array}{cc} 1 \\ 0 \\ \end{array} } \right) \ne 0$, 
these states are proportional to each other\footnote{On another basis where 
$Q_1$ and $Q_2$ are not both off-diagonal only states of the form 
$\alpha \left( {\tiny \begin{array}{cc} 1 \\ 0 \\ \end{array} } \right) + \beta \left( {\tiny \begin{array}{cc} 0 \\ 1 \\ \end{array} }
\right)  $
are invariant under a linear combination of $Q$'s. For example one may take $Q_1$ proportional to $\tau_1$ and 
$Q_2$ proportional to $\tau_2$.}.
Defining $S=\frac{1}{2} (i \epsilon Q_1+Q_2)$, one obtains 
$S^2=(S^\dagger)^2=0$ but $\{S,S^\dagger \}=H-Z$. 
This shows that one still has a two-dimensional representation for a massive non-BPS multiplet.

On the other hand, massive BPS representations are 1-dimensional, and consist of the
one state $| kink \rangle + c_0 | kink \rangle $ where $| kink \rangle$
is the kink vacuum, and $c_0$ is the fermionic zero mode. This was first mentioned 
in a footnote in ref. \cite{shifman}, and further explained in ref. \cite{ritz}
where an $N=2$ susy system was reduced to $N=(1,1)$ susy. We shall discuss
this result from a different point of view in ref.  \cite{fred}. 

Given that massive non-BPS multiplets have 2 states
and massive BPS multiplets have only one state,
multiplet shortening {\bf does} occur, and this 
explains and predicts that $M=Z$ to all loop order.

\section{Conclusions.}

We have analyzed several regularization schemes for the one-loop 
corrections $M^{(1)}$ to the mass and $Z^{(1)}$ to the central charge of the bosonic
and the susy kink. By far the simplest is in our opinion the scheme 
\cite{misha} in which
one first evaluates $\frac{\partial}{\partial m}M^{(1)}$,
where $m$ is the renormalized mass parameter,
and then 
integrates for $M^{(1)}$ by imposing the renormalization condition 
$M^{(1)}(m=0)=0$. 
This scheme should only be used for such boundary conditions
that the infinities in $M^{(1)}$ cancel. There are many such 
boundary conditions, for example periodic ones and antiperiodic ones. However,
for supersymmetric boundary conditions the bosonic and fermionic terms in the
sums cancel and one is only left with the counterterm which yields a term $m \ln (\Lambda/m)$ in $M^{(1)}$.
The sums over $\frac{\partial}{\partial m}$ derivatives of
zero point energies are only logarithmically divergent so that no ambiguities 
in combining sums arise, and the renormalization condition $M^{(1)}(m=0)=0$
seems also beyond doubt, as it states that when the two trivial vacua
($\phi=\pm m/ \sqrt{2 \lambda}$) become equal, there is no kink.
The actual computation of $M^{(1)}$ is also very simple
and yields (\ref{ura1}). This value seems now accepted at this moment by all authors 
who have recently published on this problem. The method does not depend 
on the boundary conditions for the quantum fluctuations except that $M^{(1)}$ should be finite.

We have also shown how to use momentum cut-off regularization.
A new ingredient is that one needs a smooth function $f(k-\Lambda)$ which
interpolates between the modes of the kink sector and the 
trivial sector, and we must use the same boundary conditions 
in both sectors to obtain
a finite (and correct) result.
This function $f(k-\Lambda)$ modifies the usual naive formula for the finite mass of solitons in two places: it
multiplies the naive expression for the density of states (a plausible modification which probably many 
others have considered at various times), but it also adds new term proportional $f^\prime(k-\Lambda)$.
We have not found a place in the literature where this term is proposed, and it certainly is new in the recent literature
on the kink.
The results do not depend on which boundary conditions one chooses
as long as they are the same in both sectors. If they are different,
$M^{(1)}$ may even come out infinite. 

Mode number regularization gives results which depend critically on
the boundary conditions.
We shall discuss mode regularization in more detail in another publication \cite{fred}
but we will make here some comments.
We first considered the kink-antikink system where any boundary conditions
on plane waves for the fermions 
are allowed
as far as consistency in (\ref{granferm}) is concerned. 
(The reason is that the potentials for $\psi_1$ and $\psi_2$ give the
same phase shift, see figure 2).
For the susy system, we required that the sums of modes
in the free case canceled, but one must 
choose the same boundary conditions in the $K \bar{K}$ system as
in the free sector; otherwise one creates extra positive energy
which is needed to twist the fields at the boundary. 
Given that the modes in the free sector
cancel, it follows that one needs the same boundary conditions for
bosons and fermions in the nontrivial sector of the $K\bar{K}$ system.
For the bosonic case one needs the same boundary
conditions in the trivial sector as in the nontrivial sector, 
for the same reason.
(For example,
periodic conditions in the free case and
anti-periodic conditions in the $K \bar{K}$ case, and the reverse situation, both
make $M^{(1)}$ larger). 

We then read off the corresponding boundary conditions for the
kink system. 
The result of the $K \bar{K}$ analysis for the bosonic kink is that
one gets the same 
answer if one uses periodic or anti-periodic conditions,
Dirichlet, Neumann or mixed conditions ( $\eta (-L/2)= \eta (L/2)=0$ or
$\eta (-L/2)= \eta^\prime (L/2)=0$)
in the kink sector, but then one should always use the same boundary 
conditions in the free sector (using, of course, the same 
interval as for the kink, which we took of length $L$).
For example, vanishing of $\eta$ for the $K\bar{K}$ system at the endpoints implies that $\eta$ is either symmetric or antisymmetric around the point between
the kink and antikink. Thus at this point $\eta=0$ or $\eta^\prime=0$. 
For the fermions there is the complification that the potentials for
$\psi_1$ in the kink and antikink system are different. So there is 
no natural point in between where both $\psi_1$ and $\psi_1^\prime$ vanish.
However, by averaging over all possibilities at the point exactly in the middle, one still gets the correct answer.
Two typical results thus obtained for the
boundary conditions for the
susy kink  are: 

I) For the bosons, take one-half of the sum of  $\eta(L/2)=\eta(-L/2)=0 $ 
and $\eta(L/2)=\eta^\prime(-L/2)=0$;  for the fermions take one-quarter of the
sum of  $\psi_1(L/2)=\psi_1(-L/2)=0$, $\psi_1(L/2)=\psi_1^\prime(-L/2)=0$,
$\psi_2(L/2)=\psi_2(-L/2)=0$, $\psi_2(L/2)=\psi^\prime_2(-L/2)=0$.

II) $\eta(-L/2)=\pm \eta (L/2)$, $\eta^\prime (-L/2)=\pm \eta^\prime (L/2)$ 
with either two $+$ signs or two $-$ signs ;
further $\psi_1(-L/2)=\pm \psi_2(L/2)$
and  $\psi_2 (-L/2)=\pm \psi_1 (L/2)$ with either two $+$ signs or two $-$ signs.

There are, of course, many more boundary conditions, but in
general they give incorrect results for mode regularization. In particular, 
susy boundary conditions give an incorrect result, namely $M^{(1)}=\Delta M_{susy}$
which is infinite. 
As explained in [12,6] one should remove the energy stored at the endpoints
to obtain the mass of the kink.
The zero point energies of bosons and fermions cancel each other, because the action, background and boundary conditions are all
supersymmetric. This can easily be checked.
The susy boundary conditions of \cite{schonfeld} read $\eta(L/2) = \eta(-L/2)=\psi_1(L/2)=
\psi_1(-L/2)=0$.
On the other hand, the susy boundary conditions of \cite{shifman} are given by
$\partial_x \eta \pm m \eta = \psi_2=0 $ at $\pm L/2$ (which come from $\partial_x \phi + U =0$ and imply $ (\partial_x + U^\prime) \psi_1=0 $). In each case one obtains the same 
bosonic and fermionic spectrum.

In an early paper \cite{schonfeld}, Schonfeld found 
that one can apply mode number regularization for the susy kink 
if one uses one more fermionic than bosonic mode in the continuum.
We explained this fact in two ways. On the double interval
with a kink-antikink 
one can count the
number of discrete states \cite{fred}, and one finds 4 bosonic but 3 
fermionic discrete
levels. 
We
can also explain why we need one extra fermionic mode in the kink system itself.
Consider twisted boundary conditions for the fermions (
 $\psi_1(-L/2)=\pm \psi_2(L/2)$
and  $\psi_2 (-L/2)=\pm \psi_1 (L/2)$ either with two $+$ signs or
with two $-$ signs ). 
We consider these boundary conditions because in ref \cite{misha}
it was argued that these conditions create no
extra energy at the boundary. We can now prove this claim.
Extending the single interval of length $L$ to a double
interval, 
one obtains periodic 
boundary conditions 
(for example $\psi_1(-L) \to - \psi_2(0)$ and then $- \psi_2(0) 
\to \psi_1(L)$)
which indeed have no extra energy at the boundaries.
This explains from a geometrical point why the twisted 
boundary conditions for the supersymmetric kink are consistent for plane waves. In \cite{misha}
a direct algebraic proof of their consistency was given
and the notion of M\"obius strip was introduced. The double interval with
periodic boundary conditions can indeed be viewed as a M\"obius strip. On this strip
bosons have 4 discrete states but fermions only 3. 

We conclude that all regularization schemes used in the past few years to
compute $M^{(1)}$ are now giving the same results, so we consider the
problem of computing $M^{(1)}$ as solved.

We also provided a new scheme with two cut-off parameters to compute
$Z^{(1)}$. We noted that higher derivative regularization \cite{shifman}
by itself does not 
regulate the one-loop graphs which yield $Z^{(1)}$. We also
noted that the phase shift analysis in the form
presented in [3,4] implicitly assumes dimensional regularization.

Although the result that $M=Z$ holds to all loop orders due to multiplet
shortening \cite{shifman}, it would be interesting to
study whether there are nonperturbative corrections
to the kink mass, and whether susy is dynamically broken.
Similar questions can now be attacked in higher dimensional models.
\\ \\

{\bf Acknowledgments: }It is a pleasure to thank A.Goldhaber, N.Graham, A.Rebhan, M.Stephanov and C.Vafa for discussions
and E.Fahri and R.L.Jaffe for correspondence. 
In particular we thank M.Shifman, A.Vainshtein and M.Voloshin for detailed clarifying discussions and correspondence.
We found it very interesting to speak to one of the pioneers of the field, J.Schonfeld, and discover in this
way some background motivations for early work on susy BPS states.

\end{document}